\documentclass[fleqn,usenatbib]{mnras}

\usepackage{newtxtext,newtxmath}
\usepackage[T1]{fontenc}
\usepackage{ae,aecompl}

\usepackage{graphicx}	% Including figure files
\usepackage{amsmath}	% Advanced maths commands
\usepackage{amssymb}	% Extra maths symbols

\title[Slingshot prominences]{Slingshot prominences: coronal structure, mass loss and spin down}

\author[M. Jardine et al.]{
M. Jardine,$^{1}$\thanks{E-mail: mmj@st-andrews.ac.uk}
A. Collier Cameron,$^{1}$
J.-F. Donati$^{2}$
and G.A.J. Hussain$^{3}$
\\
$^{1}$SUPA, School of Physics and Astronomy, North Haugh, St Andrews, Fife, KY16 9SS, UK\\
$^{2}$Laboratoire d'Astrophysique, Observatoire Midi-Pyr\'en\'es, 14 Av. E. Belin, F-31400 Toulouse, France\\
$^{3}$European Southern Observatory, Karl-Schwarzschild-Str. 2, D-85748 Garching bei M\"unchen, Germany;\\ Institut de Recherche en Astrophysique et Plan\'etologie, Universit\'e de Toulouse, UPS-OMP, F-31400 Toulouse, France
}

\date{Accepted XXX. Received YYY; in original form ZZZ}

\pubyear{2018}

\begin{document}
\label{firstpage}
\pagerange{\pageref{firstpage}--\pageref{lastpage}}
\maketitle

\begin{abstract}
The structure of a star's coronal magnetic field is a fundamental property that governs the high-energy emission from the hot coronal gas and the loss of mass and angular momentum in the stellar wind. It is, however, extremely difficult to measure.  We report a new method to trace this structure in rapidly-rotating young convective stars, using the cool gas trapped on coronal field lines as markers. This gas forms ``slingshot prominences'' which appear as transient absorption features in H-$\alpha$. By using different methods of extrapolating this field from the surface measurements, we determine locations for prominence support and produce synthetic H-$\alpha$ stacked spectra. The absorption features produced with a {\it potential} field extrapolation match well this those observed, while those from a {\it non-potential} field do not. In systems where the rotation and magnetic axes are well aligned, up to $50\%$ of the prominence mass may transit the star and so produces a observable feature. This fraction may fall as low as $~2\%$ in very highly inclined systems. Ejected prominences carry away mass and angular momentum at rates that vary by two orders of magnitude, but which may approach those carried by the stellar wind.

\end{abstract}

\begin{keywords}
stars: coronae, stars: late-types, stars: magnetic fields, stars: rotation, stars: solar-type
\end{keywords}

%%%%%%%%%%%%%%%%%%%%%%%%%%%%%%%%%%%%%%%%%%%%%%%%%%

%%%%%%%%%%%%%%%%% BODY OF PAPER %%%%%%%%%%%%%%%%%%

% SECTION 
%--------------------------------------------------------
\section{Introduction}

The structuring of the solar corona into magnetically-confined X-ray bright loops and dark, wind-bearing coronal holes has been studied for many decades. Over the course of the solar cycle, this structure evolves, leading to variations not only in the Sun's X-ray luminosity, but also the mass and angular momentum lost in the solar wind. This structure has also evolved as the Sun has been spun down by the loss of angular momentum in its wind \citep{2007LRSP....4....3G,2014MNRAS.441.2361V}.

The nature of the corresponding structures in other solar-like stars is difficult to determine, however, without resolved observations.   In binary systems, X-ray eclipse observations can provide information about the location of emitting structures \citep{1996ApJ...473..470S,2001A&A...365L.344G}. These early studies showed highly structured coronae, with localised regions of emission consistent with Doppler imaging results. High time cadence, high resolution studies of X-ray spectra can also use the velocity shifts of X-ray lines to localise the emission in velocity space, giving information about the extent of the confined X-ray corona \citep{2007MNRAS.377.1488H}.  More recently, the possibility of exploiting exoplanetary transits to probe the structure of the exoplanetary exosphere has generated new interest in the structure and variability of the underlying stellar emission \citep{2015ApJ...802...41L}. Observational studies of the stellar winds that correspond to these X-ray coronae are hampered by the wind's low density. The thermal radio emission from the wind can be used to provide measurements of the wind density \citep{1975A&A....39....1P} but these are typically upper limits \citep{1996ApJ...462L..91L,1997A&A...319..578V,2000GeoRL..27..501G,2014ApJ...788..112V}. More recently,  \citet{2017A&A...599A.127F} have provided stringent upper limits on mass loss rates for four solar-type stars based on a range of optical depth regimes. An alternative method is the novel technique that has been developed using the enhanced Lyman-$\alpha$ absorption in the ``hydrogen wall'' that forms at the stellar asterosphere. This provides a probe of the wind density and hence, assuming a simple wind model, the mass loss rate \citep{2004LRSP....1....2W}. The results suggest that mass loss rates increase with X-ray flux up to some ``wind dividing line'' but beyond that they appear to decrease. 

The coronal magnetic field that produces this structure remains elusive. It is only at stellar surfaces that we can detect and measure the magnetic field. Its geometry can be revealed by spectropolarimetric studies, using the technique of Zeeman-Doppler imaging \citep{1997MNRAS.291....1D,1999MNRAS.302..437D,2000MNRAS.318..961H,2012A&A...548A..95C,2015ApJ...805..169R}. This produces maps of the vector components of the surface magnetic field, often decomposed into their toroidal and poloidal components \citep{2006MNRAS.370..629D}. The presence of toroidal fields at the surfaces of stars was initially surprising because it is different from what is found on the Sun \citep{1992A&A...265..682D,1999MNRAS.302..457D,2003MNRAS.345.1145D}, but they have now been detected in a range of stars including the very young T-Tauri stars \citep{,2010MNRAS.403..159S,2014MNRAS.444.3220D}. While toroidal fields are detected on a range of stars, it is in stars with a tachocline that they may contribute a significant fraction of the total magnetic energy \citep{2008sf2a.conf..523P,2015MNRAS.453.4301S}. If this toroidal field extends beyond the surface into the corona, it may have a significant effect on the coronal structure and dynamics. \citet{2018ApJ...862...93A} suggest that it may act to enhance the confinement of coronal plasma, inhibiting the ejection of CMEs on very active stars. This may explain why the solar scaling of CME kinetic energy with X-ray flux cannot be extrapolated to active stars without requiring an unphysically large energy for stellar CMEs \citep{2013ApJ...764..170D}. The development of a strong toroidal field that may inhibit the stellar wind has also been proposed as an explanation for the apparent drop in mass loss rate per unit surface area beyond the ``wind dividing line''  \citep{2004LRSP....1....2W}. As shown by \citet{2016MNRAS.455L..52V}, however, there is no apparent change in magnetic field topology across this line. 

%-------------------Figure 1 ----------------------
%-------------------Dipole cartoon ----------------------
% Figure 1 ----------------------------------------
\begin{figure}
\begin{centering}
    \includegraphics[width=5cm]{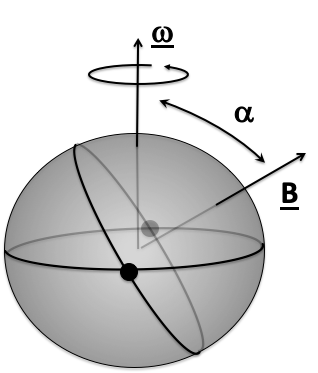}
    \caption{Schematic view of the inclination of the dipole axis to the rotation axis. The rotational and magnetic equators intersect at two longitudes, shown by black circles.}
    \label{fig:dipole_cartoon}
\end{centering}
\end{figure}
% End figure ----------------------------------------
%--------------------------------------------------------

In order to determine the nature of the stellar coronae that correspond to the observed surface magnetic fields, we need to understand how these fields are extrapolated into the corona. This requires some observational signature of the height and geometry of the magnetic loops that form the corona. One of the most successful methods of localising emission is to use an eclipse observation, but for a single star without a binary companion or a transiting planet, it might appear that there is no object to provide an ``occulting disk''. Many young, rapidly-rotating stars exhibit transient H-$\alpha$ absorption features, however, due to the centrifugal trapping of cool, dense gas in ``slingshot prominences'' \citep{collier1989I,collier1989II}. These have been detected in both single and binary stars \citep{Barnes2000,Barnes2001,Byrne1996,collier1992,Dunstone2006,Eibe1998,Hall1992,Skelly2008,Skelly2009,Petit2005}. These ``slingshot prominences'' co-rotate with the star, scattering H-$\alpha$ photons out of the line of sight when they pass between the observer and the stellar disk. The rotation phase and drift rate of these absorption features travelling through the absorption line allows us to locate them within the stellar corona. They mark locations not only of closed loops, but regions within the coronal magnetic field where stable equilibria are possible. While they generally appear as absorption features, in a few systems where the co-rotation radius is very close to the stellar surface, they are also seen in emission \citep{Donati2000,2006MNRAS.373.1308D,Kolbin2017}. Their relation to the photometric "dips" seen in some K2  lightcurves of M dwarfs is as yet unknown \citep{Stauffer2017}.

The presence of these cool clouds trapped in the coronae of rapidly-rotating stars is observed sufficiently frequently that they must be a common feature of stars where centrifugal support within a corona is possible. The early simple models of slingshot prominences supported this idea. Modelling of the mechanical support of these prominences in both single  \citep{1995A&A...298..172F, 1996A&A...305..265F,2000MNRAS.316..647F} and binary systems \citep{1998A&A...335..248F} demonstrated that the observed surface field strengths are adequate to support the derived prominence masses, and that these would be expected to cluster around the Keplerian co-rotation radius. Models of prominence thermal equilibrium \citep{1997A&A...327..252F} showed that cool condensations within hot loops of size equal to a few stellar radii can easily be produced without the need for special heating functions, while sequences of mechanical equilibria at a range of temperatures and surface pressures exist for simple background field structures  \citep{2005MNRAS.361.1173J,2019MNRAS.483.1513W}.

The existence of a series of cool equilibria does not of course guarantee that there is a dynamical path to access them. A thermal instability in a loop summit may cause a drop in pressure there which will drive an upflow from the surface. This upflow may continue until pressure balance is restored, and some new, cooler equilibrium is found. If the upflow becomes supersonic, however, before it reaches the condensation at the loop summit, the surface will not be able to respond and will continue to drive a hot upflow into the loop summit. The accumulation of mass there will eventually exceed the ability of the magnetic field to support it, and the mass will either fall back towards the surface (if it has condensed below the co-rotation radius) or will be expelled if it has formed above the co-rotation radius. This ``limit-cycle'' behaviour will continue, effectively acting as an intermittent form of stellar wind. The criterion for this to occur is that the loop temperature exceeds a critical value 
\begin{equation}
T_{\rm crit} [10^6K] = 1.6 \left(\frac{M_\star [M_\odot]}{P [d]} \right)^{2/3}.
\label{Tcrit}
\end{equation}
\citet{2019MNRAS.482.2853J} showed that a significant number of low mass stars lie in this regime and that the observed masses and lifetimes of prominences may therefore be used as a measure of the rate of mass upflow and hence of the stellar wind mass loss rate. These values for stars with high X-ray fluxes (and hence high coronal temperatures) show that wind mass loss rates continue to increase with increasing X-ray flux, well beyond the point where previous measurements had suggested they might saturate \citep{2004LRSP....1....2W}. 

This apparent saturation can be attributed to a large scatter in mass loss rates, coupled with a small number of measurements. What is not clear, however, is whether this scatter is intrinsic or not. It may be that each star has a  mass loss rate that varies in time, due perhaps to changes during a magnetic cycle. For each star, however, we typically have one observation, often made at one viewing angle. If the wind is spatially variable, this may lead to a large variation.

The nature of the winds from these stars is therefore dependant on the structure of the corona, and how this might vary from one observing epoch to the next. In order to investigate this, we consider the rapidly-rotating (P$_{\rm rot} = 0.514$d) active K0 dwarf AB Dor, on which slingshot prominences were first discovered. Zeeman-Doppler surface magnetograms and simultaneous H-$\alpha$ spectra have been acquired for this star almost annually from 1995-2007. This provides a rich history of the year-to-year variations in the magnetic structure and prominence locations for this star, allowing us to assess the role of the strong surface toroidal field observed on this star and the variations in the prominence masses supported.

% SECTION 
%--------------------------------------------------------
\section{Prominence support within a dipole field}

Our approach is to model first the 3D structure of the coronal magnetic field, and then to determine the distribution of cool material that could be trapped within it. We assume that the field is strong enough that we may neglect any distortion produced by the prominence (see \citet{2019MNRAS.483.1513W}). From this mass distribution, we then predict the absorption transients that would be produced in H-$\alpha$ and compare these with the observed stacked H-$\alpha$ spectra. In order to develop our understanding, however, we first begin by assuming that the field is a dipole inclined at some angle $\alpha$ to the rotation axis (see Fig. \ref{fig:dipole_cartoon}). 

%--------------------Figure 2----------------------
%-------------------Dipole field plus Halpha----------------------
% Figure 2 ---------from allsp3_stable_plus_wind.pro---------
\begin{figure*}
\begin{centering}
    \includegraphics[width=8.cm]{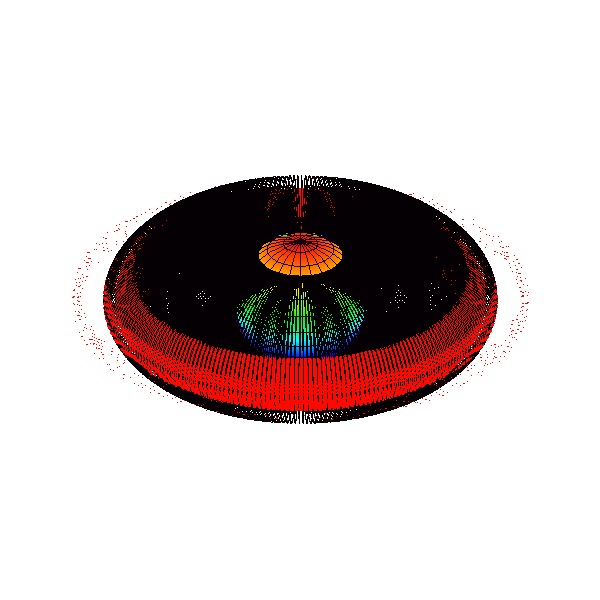}
    \includegraphics[width=5.cm]{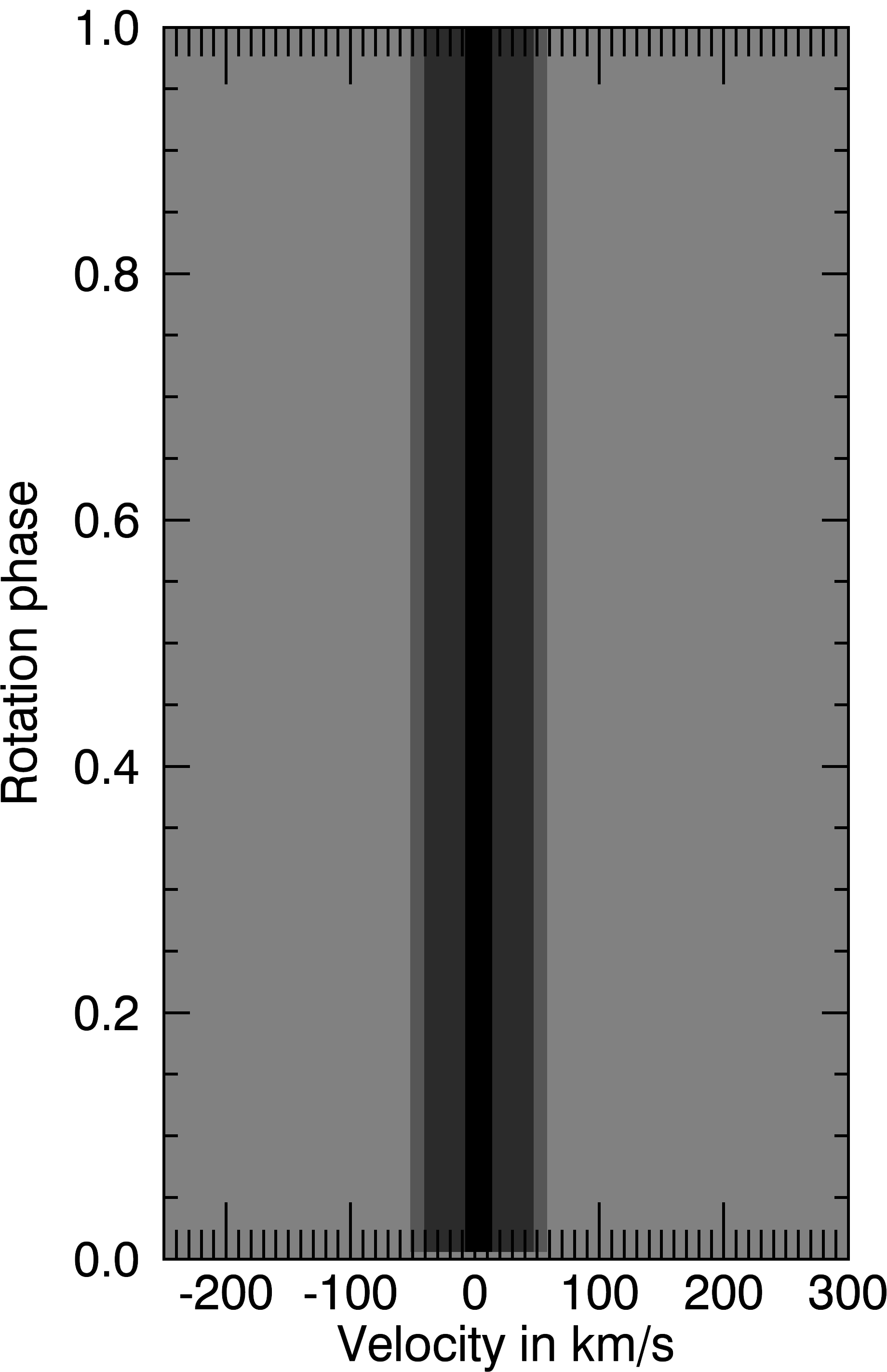}\\
    \includegraphics[width=8.cm]{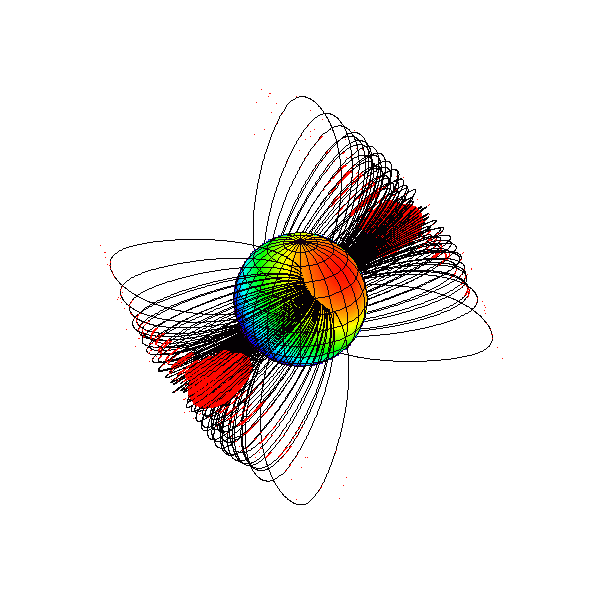}
    \includegraphics[width=5.cm]{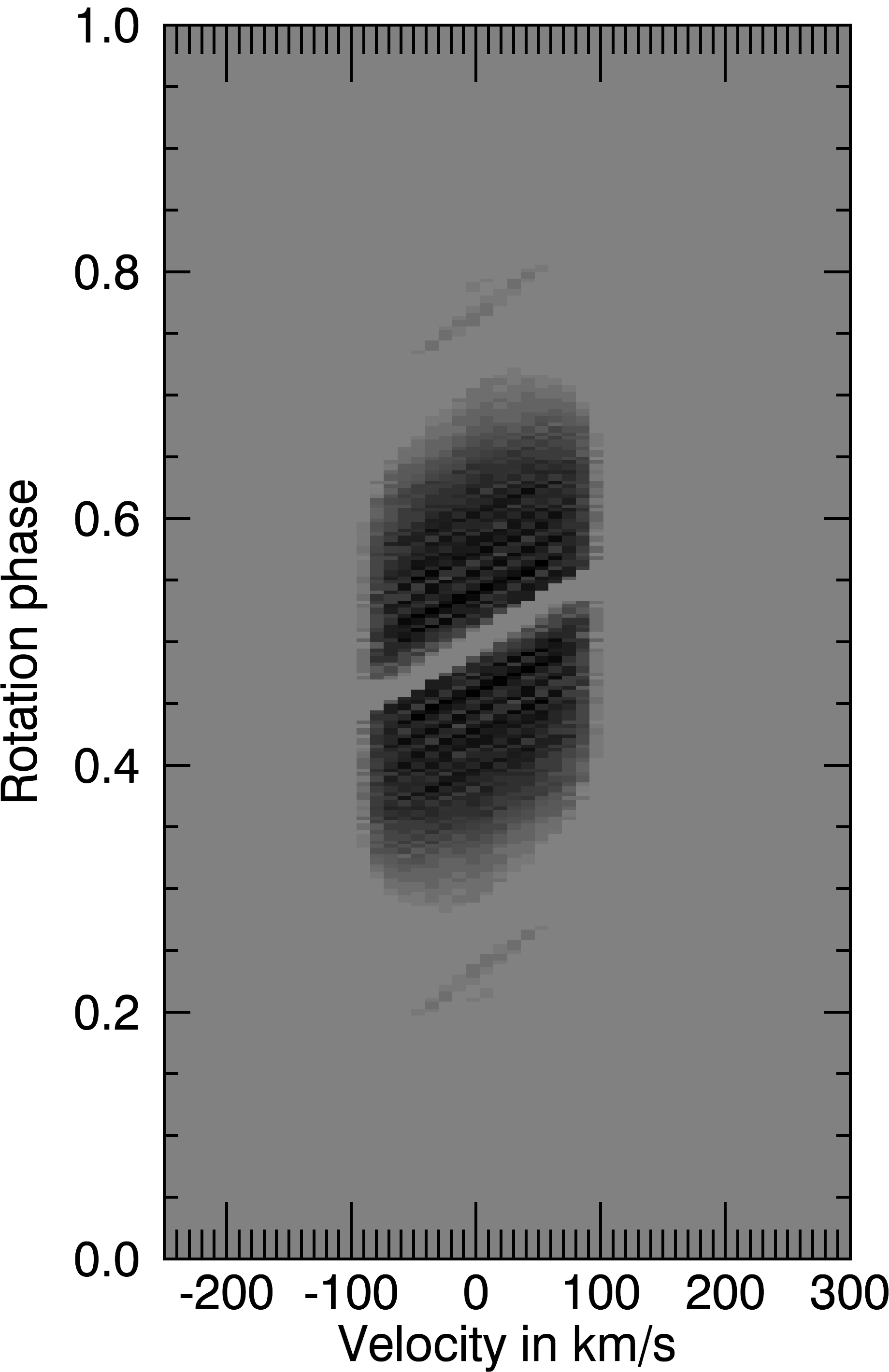}\\
    \includegraphics[width=8.cm]{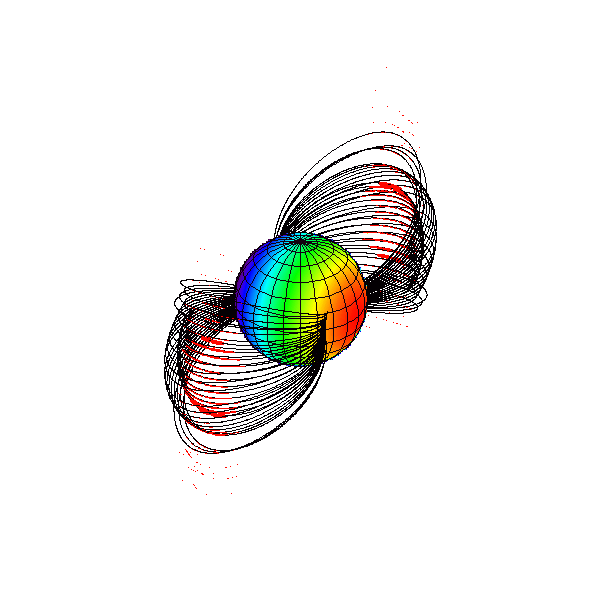}
    \includegraphics[width=5.cm]{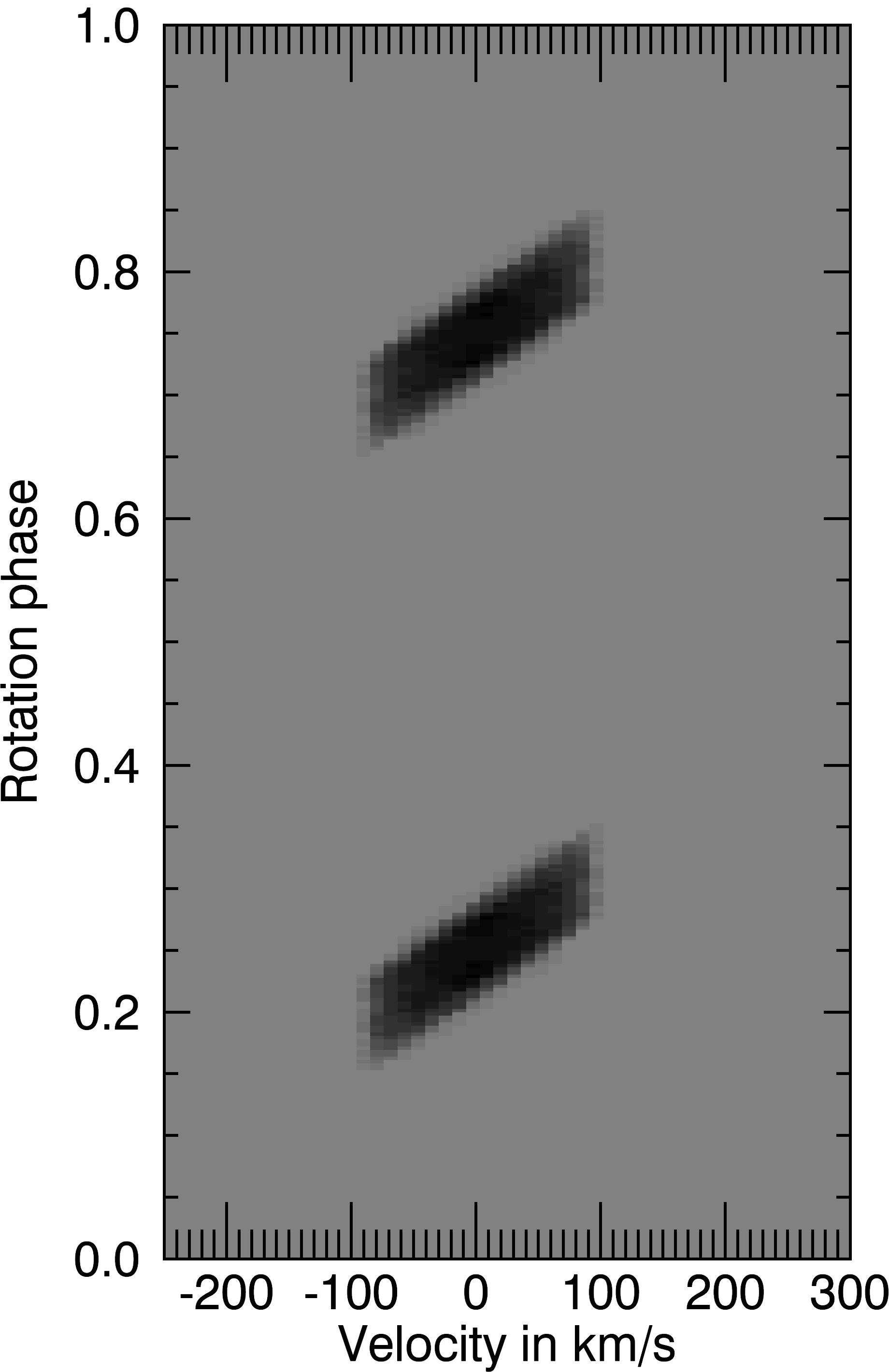}
    \caption {The left column shows in red the cool prominences and in black a selection of the magnetic field lines that support the most massive prominences, while the right column shows the corresponding H-$\alpha$ transients. The top row shows a dipole inclination to the rotation axis of 0$^\circ$, the middle row 30$^\circ$ and the bottom row 90$^\circ$ respectively.}
    \label{fig:dipole_pix}
\end{centering}
\end{figure*}

% End figure ----------------------------------------
%--------------------------------------------------------

% SUBSECTION 
%--------------------------------------------------------
\subsection{Predicting prominence locations}
\label{subsection:prom_model}

Within this field structure, we determine the locations where cool gas may be supported in a stable equilibrium \citep{2000MNRAS.316..647F}. A particle placed at such a point will be in a gravitational potential minimum {\it as calculated following the direction of the field} \citep{2001MNRAS.324..201J}. This requires that for such a point,
\begin{equation}
\left( \underline{B} \cdot \underline{\nabla} \right) \left(  \underline{g}_{\rm eff} \cdot \underline{B} \right) < 0
\end{equation}
where the effective gravitational acceleration is given by 
%$g_{\rm eff} =( {\bf g.B})/|{\bf B}|$ and 
\begin{equation}
\underline{g}_{\rm eff}(r,\theta) = \left( -GM_{\star}/r^{2} + 
                     \omega_\star^{2}r\sin^{2}\theta,
		     \omega_\star^{2}r\sin\theta\cos\theta 
             \right), 
\end{equation}
where $\omega_\star$ is the stellar angular velocity. We set any field line that passes through such a point to have a temperature of 8500K characteristic of stellar prominences \citep{1990MNRAS.247..415C} and assume that the plasma density at the stable point is given by its maximum value of
\begin{equation}
\rho_{\rm max} = \frac{B^2}{\mu R_c |\underline{g}_{\rm eff}|}
\label{eq:rho_max}
\end{equation}
where $R_c$ is the local radius of curvature \citep{2018MNRAS.475L..25V}. Using this as a boundary condition, we can then calculate the corresponding hydrostatic distribution of mass using the known flux tube volume. The mass that can be supported therefore varies as $B^2$ which is a measure of the magnetic energy per unit volume. All of our dipole models have the same field strength however and so only the field geometry is varied. The other very strong dependence that can be seen in equation (\ref{eq:rho_max}) is on the local gravitational acceleration $g_{\rm eff}$. Close to the co-rotation radius, where $g_{\rm eff} \rightarrow 0$, the maximum mass that can be supported is largest. Field lines that pass close to this point are therefore able to support most mass. Fig.~\ref{fig:dipole_pix} shows the hydrostatic distribution of this cool material within the corona. The clustering of prominence material close to the co-rotation radius is apparent. 

%-------------------Figure 3-----------------------
%------------------- Masses - dipole -----------------------
% Figure 3 ----From (plot)prom_mass_time.pro and plotprom_stats.pro--
\begin{figure}
\begin{centering}
    \includegraphics[width=\columnwidth]{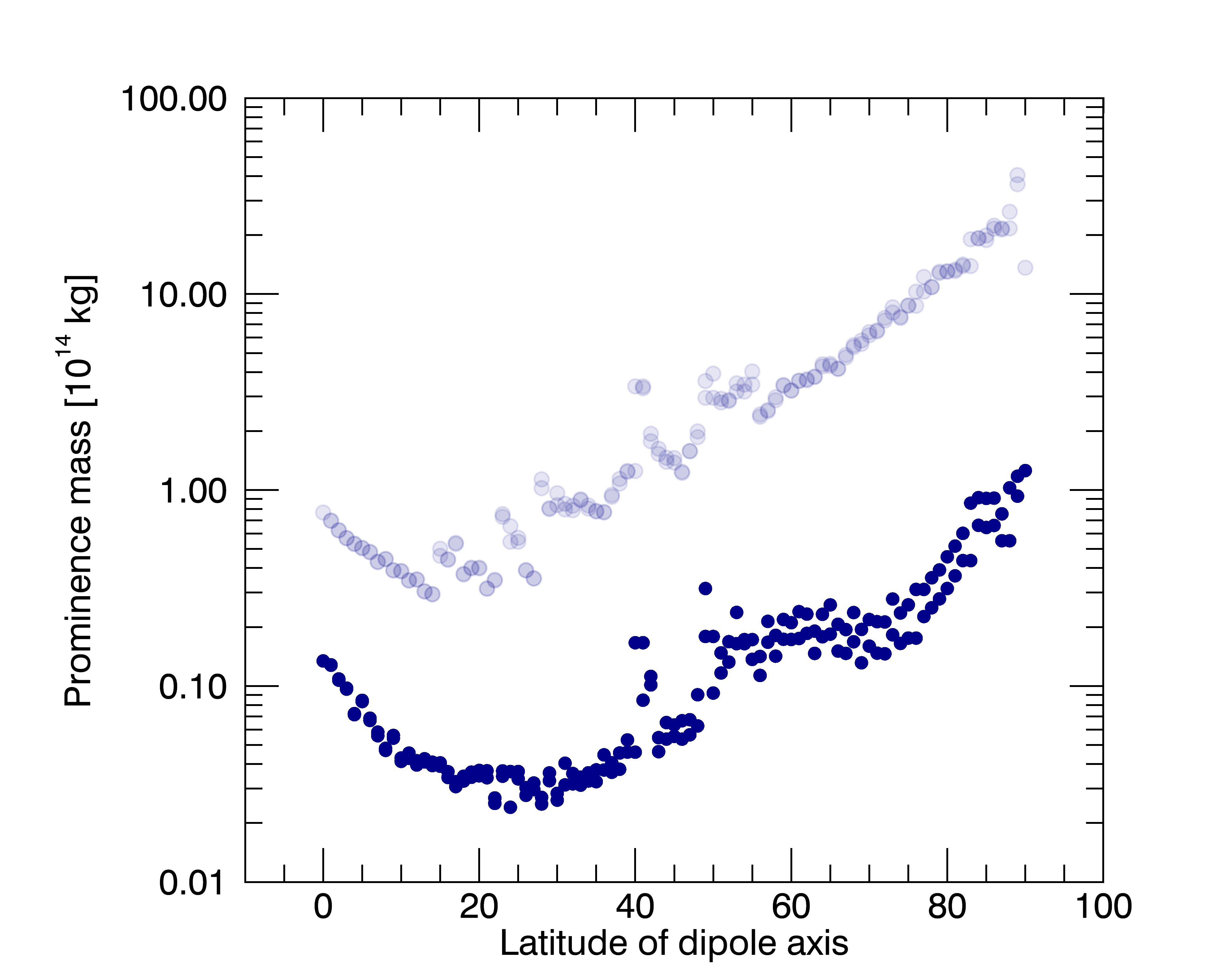}
    \caption{The influence of the latitude of the dipole axis on the mass of prominences that can be supported. The faint (upper) symbols show the total mass and the dark (lower) symbols show the mass that is visible, assuming that the rotation axis is inclined at 60$^\circ$ degrees to the line of sight. The dipole field strength is set to 40G which is the average value of the dipole component of AB Dor's field over 1995-2007.}
\label{fig:dipole_mass_lat}
\end{centering}
\end{figure}
% End figure ----------------------------------------
%--------------------------------------------------------

When the magnetic axis is aligned with the rotation axis, the cool gas settles in a torus in the equatorial plane. Stable equilibria exist in the equatorial plane for radii $r > 0.87 r_\star$ (see, for example, \citet{2000MNRAS.316..647F}). As the magnetic axis is tilted, however, this torus of magnetic loops also inclines. Since the centrifugal support is greatest in the equatorial plane, only the parts of this torus that lie close to the equator can support material. As a result, only the sections of the torus that cross the equator can be filled with prominences. The total mass that can be supported therefore decreases by a factor for 100 as the dipole latitude decreases. This can be seen clearly in Fig. ~\ref{fig:dipole_mass_lat}. Only a small fraction (typically less than $10\%$) of the mass supported can be observed as a transient absorption feature, however, as most of the cool material does not transit the stellar disk. As the latitude of the dipole axis decreases, the magnetic equator runs almost north-south and hence although less mass can be supported, a larger fraction of that mass is visible.

%---------------------Figure 4-----------------------
%--------------------- Potential field -----------------------
% Figure 4 --From plotprom_vels.pro and allsp3_stable_plus_wind.pro ----
\begin{figure*}
    \includegraphics[width=5cm]{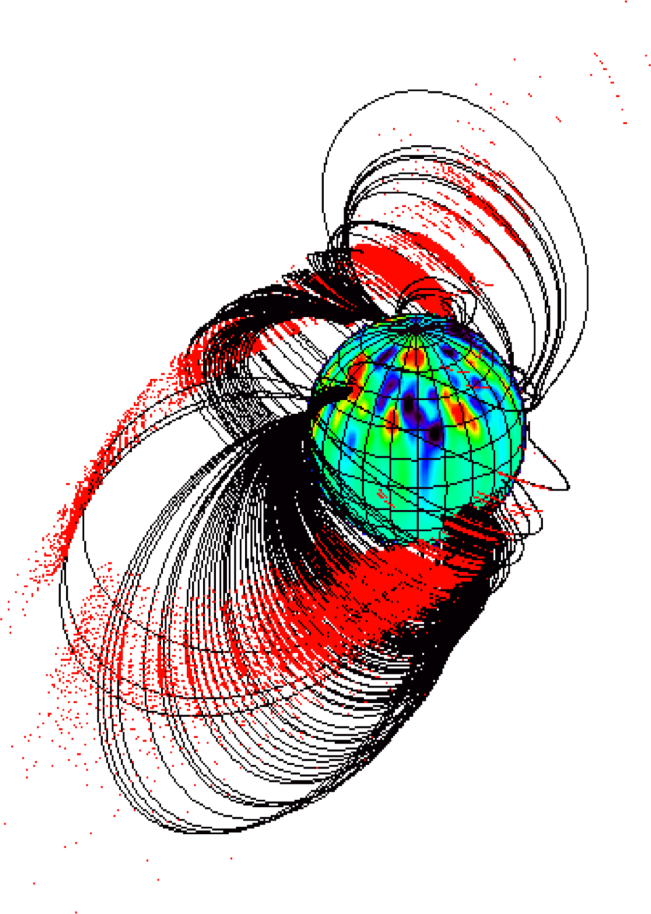}
        \includegraphics[width=5cm]{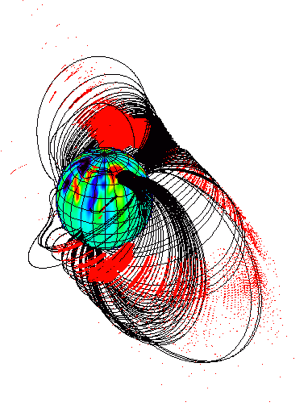}
          \includegraphics[width=3.6cm]{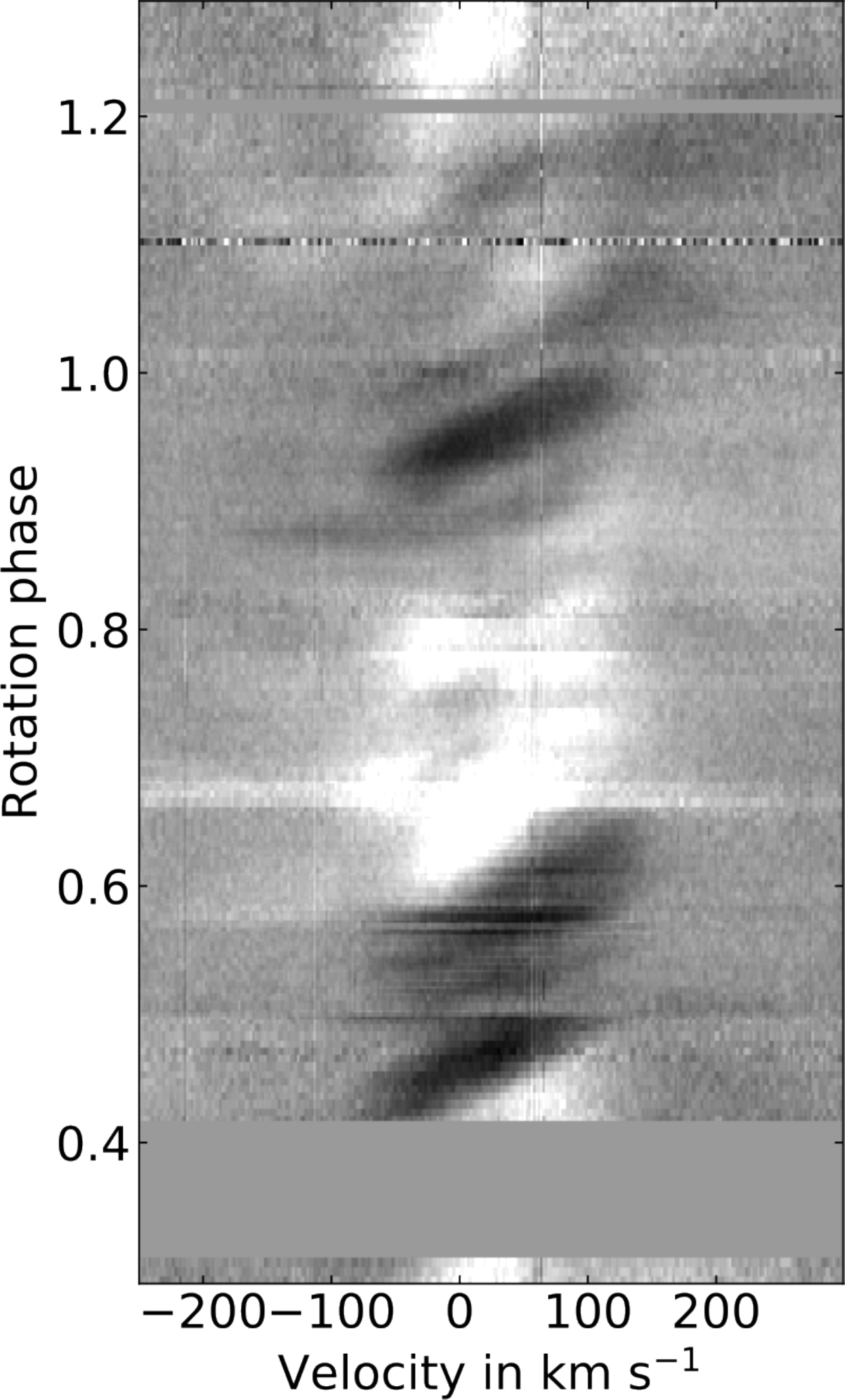}
        \includegraphics[width=3.735cm]{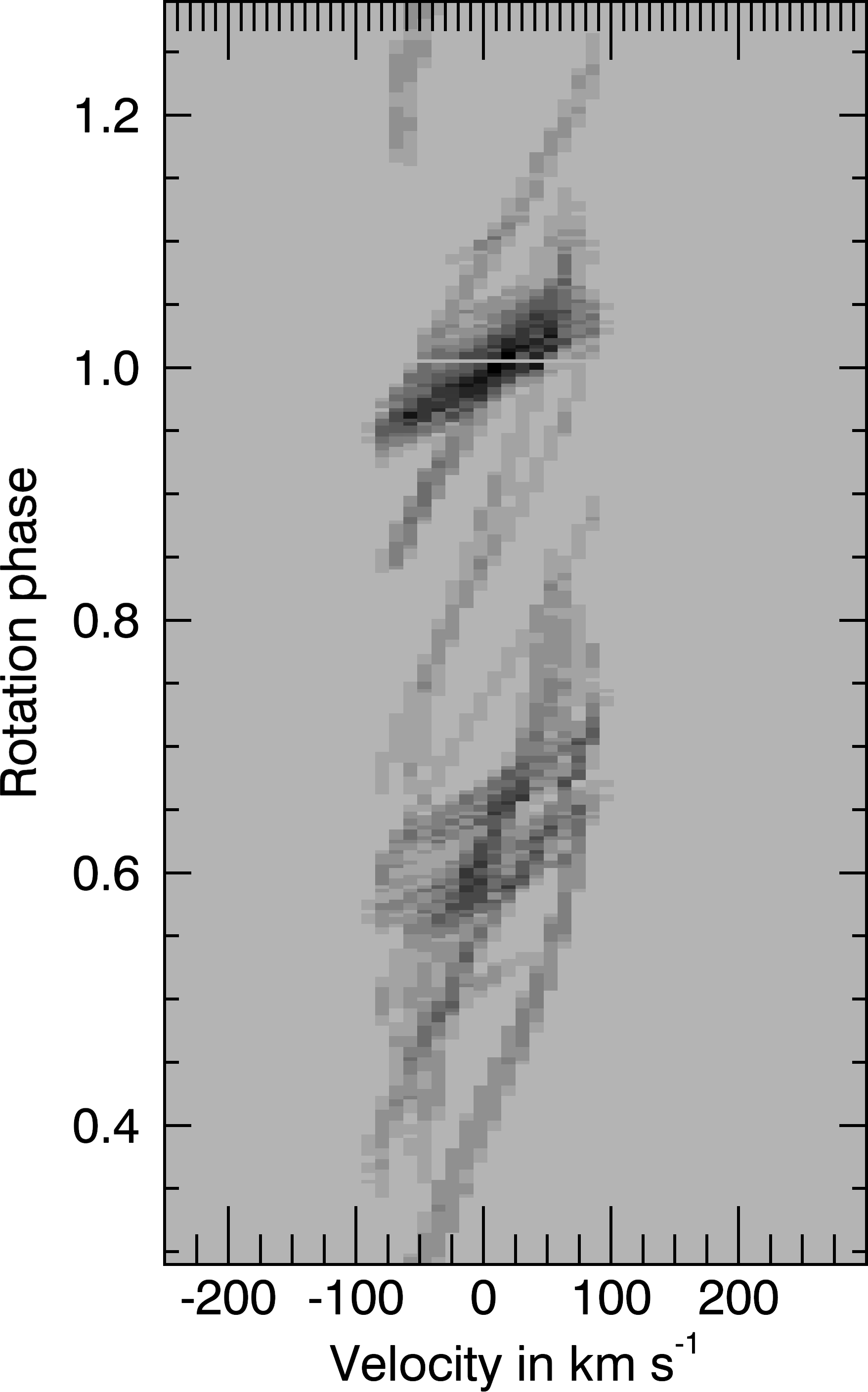}

    \caption{Predicted prominence locations and H$\alpha$ dynamic spectra for AB Dor in Dec 1996. From left to right, panels 1 and 2 show the large-scale field structure with a map of the radial magnetic field painted on the stellar surface (blue denotes negative field, while red is positive). Panel 1 shows phase 0.6, while panel 2 shows phase 1.0. The cool prominence material is show in red, and samples of the field lines supporting it are drawn in black. The third and fourth panels show the corresponding observed and synthetic H$\alpha$ dynamic spectra respectively.}
    \label{fig:1996}
\end{figure*}
% End figure ----------------------------------------
%--------------------------------------------------------
% SUBSECTION 
%--------------------------------------------------------
 \subsection{Synthetic absorption transients}
 \label{sec:synthtrans}

This cool prominence material will scatter photospheric H-${\alpha}$ photons out of the line of sight. In order to model the nature of the H-$\alpha$ absorption transients, we first calculate the optical depth along the line of sight assuming a uniform opacity for each prominence. The prominences co-rotate with the stellar magnetic field and so each element location $(r,\theta,\phi)$ has a line of sight velocity
\begin{equation}
v_{\rm los} =  \omega_\star r \sin\theta\sin i \sin(\phi - \phi_0) 
\label{eq:2}
\end{equation}
where $i$ is the inclination of the stellar rotation axis to the line of sight and $\phi_0$ is the longitude of the observer. Absorption features appear in the line profile shifted by this velocity. At line centre (where $\phi = \phi_0$) the drift rate $\dot{v}_{\rm los}$ of these features is a direct measure of the distance of the absorbing feature from the rotation axis, since here
\begin{equation}
\dot{v}_{\rm los} =  \omega_\star^2 r \sin\theta\sin i.
\end{equation}
For AB Dor, these features typically have a slope that places them at, or just beyond, the co-rotation radius. 
%We note that the centre of mass of the AB Dor A binary is receding from the solar-system barycentre at a velocity of 31.4 km~s$^{-1}$ [CITE]. 

We show in Fig. \ref{fig:dipole_pix} the synthetic stacked H-$\alpha$ spectra that correspond to each dipole inclination. If the dipole and rotational axes are aligned, prominences will form in an equatorial torus that surrounds the star. In this case, for an observer viewing the system at sufficiently high inclinations, prominences would always be in view and the H-$\alpha$ absorption features would be seen at all rotation phases if all of the available support sites are filled with prominence material. We have assumed that this is the case and so as shown in Fig.~\ref{fig:dipole_pix} there are no phases clear of absorption. As the inclination of the dipole axis is increased, gaps appear in the distribution of mass with longitude. These gaps can be seen clearly in the H-$\alpha$ spectra shown in Fig.~\ref{fig:dipole_pix}. Since the two clumps of prominence material cluster around the intersection of the magnetic and rotational equators, their associated absorption transients appear symmetrically placed about the phase of the magnetic axis. In this case, it is located at phase 0.5. The slope of the absorption transients confirms that the absorbing features are supported at the co-rotation radius.

% SECTION 
%--------------------------------------------------------
\section{Using the observed magnetograms}

The simple example of a dipole field demonstrates that the geometry of the magnetic field can strongly influence not only on the prominence mass that can be supported or detected, but also the morphology of the resulting stacked H-$\alpha$ spectra. This suggests that the differences apparent in AB Dor's magnetic field from magnetograms constructed  between 1995 and 2007 may lead to similar variations in the predicted prominence distribution. 

We therefore use these magnetograms to extrapolate the coronal magnetic field at each epoch and use the same method as in sections ~\ref{subsection:prom_model} and ~\ref{sec:synthtrans} to determine the distribution of prominence material and the resulting synthetic H-$\alpha$ spectra. 

%-------------------- Figure 5 -----------------------
% Figure 5 ----From (plot)prom_mass_time.pro and plotprom_stats.pro--
\begin{figure}
\begin{centering}
    \includegraphics[width=\columnwidth]{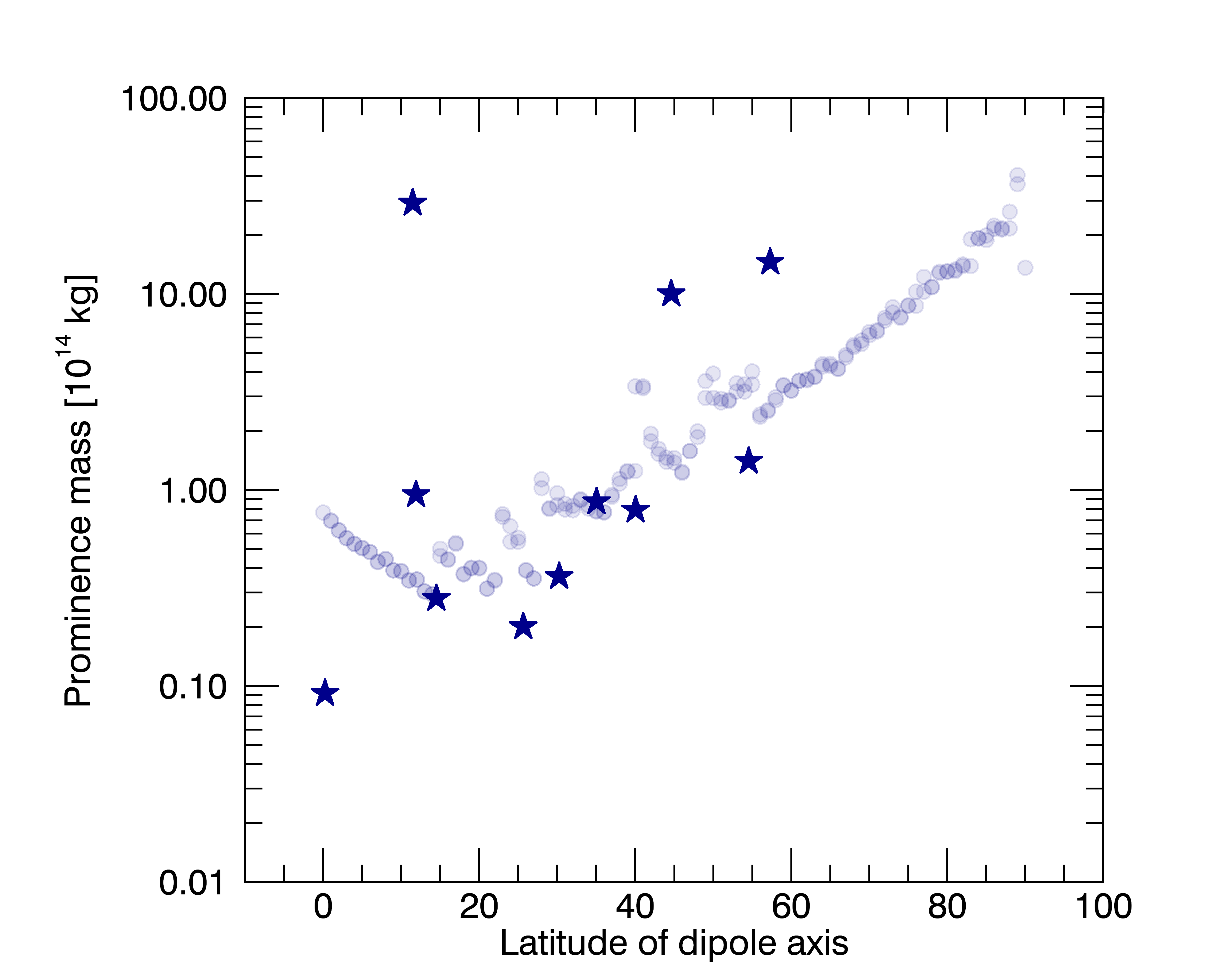}
    \includegraphics[width=\columnwidth]{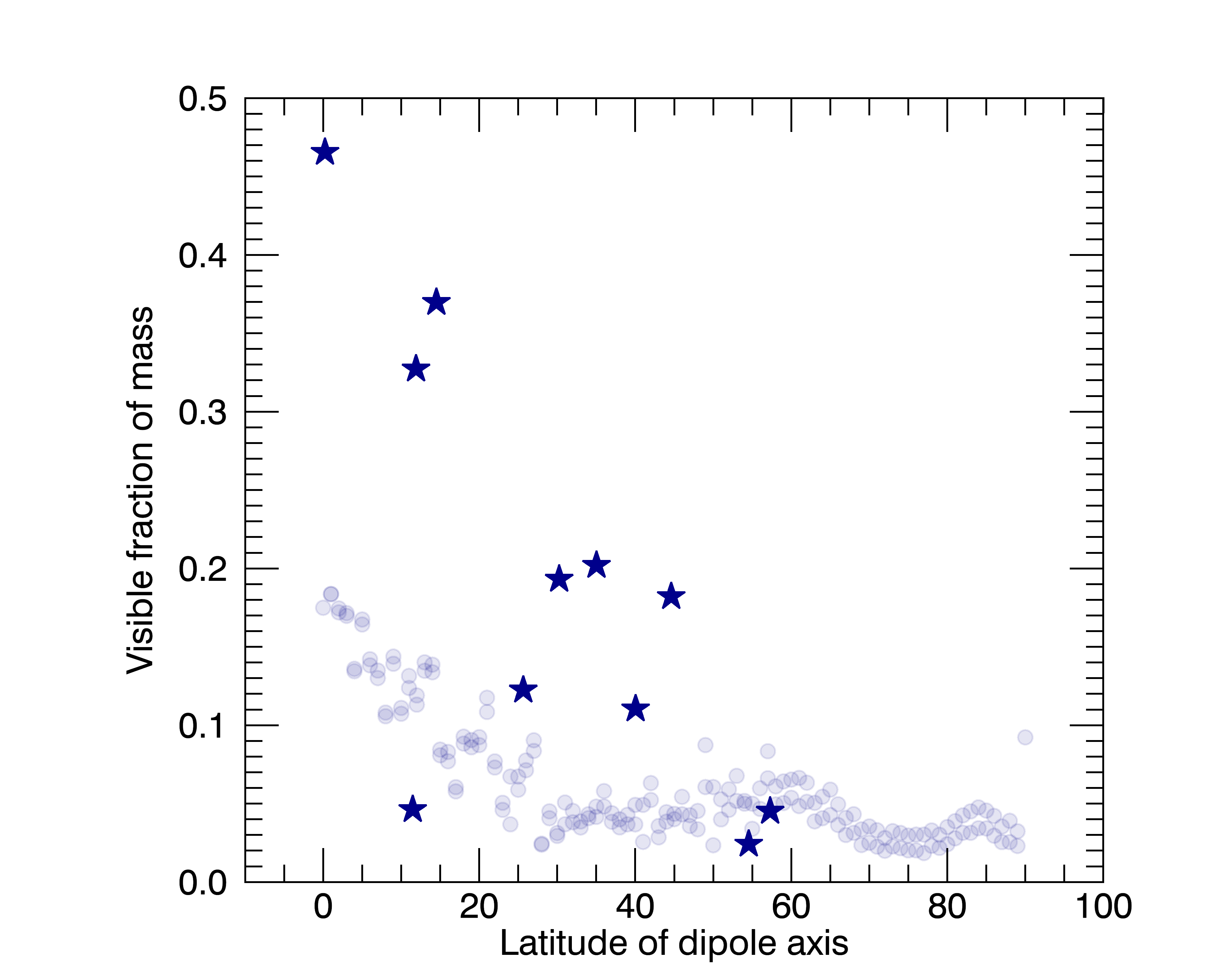}
    \caption{The variation with latitude of the dipole axis of (top) the mass of prominences and (bottom) the fraction of that mass that is visible. Values for AB Dor as shown as dark blue stars, while the faint blue circles show the values obtained for a $40$G dipole at various inclinations. In the case of AB Dor, values are shown as a function of the latitude of the dipole component of the total field.}
    \label{fig:stars_mprom}
\end{centering}
\end{figure}
% End figure ----------------------------------------
%--------------------------------------------------------

% SUBSECTION 
%--------------------------------------------------------
\subsection{A potential field}

We begin by using the {\it Potential field Source Surface} method \citep{1969SoPh....9..131A}. Since the field is assumed to be potential we may express it as $\underline{B}^{\rm pot} =-\underline{\nabla} \Psi$, where $\underline{\nabla} \cdot \underline{B}^{\rm pot} = 0$ requires $\nabla^2 \Psi = 0$. We may therefore express $\Psi$ 
in spherical co-ordinates $(r,\theta,\phi)$ as
\begin{equation}
 \Psi = \sum_{l=1}^{N}\sum_{m=-l}^{l} [a_{lm}r^l + b_{lm}r^{-(l+1)}]
         P_{lm}(\theta) e^{i m \phi},
\end{equation}
where all radii are scaled to the stellar radius. We assume that at some radius (know as the source surface, r$_{ss}$) the field is opened up by the pressure of the hot coronal gas, and so at $r=r_{ss}$, $B_\theta^{\rm pot}=B_\phi^{\rm pot} = 0$. Hence 
\begin{equation}
b_{lm}=-a_{lm} r_{ss}^{2l+1}
\end{equation}
and we may write
\begin{equation}
B_r ^{\rm pot}=  \sum^N_{l=1}\sum^l_{m=-l} 
                    B_{lm}P_{lm}(\theta)f_l(r,r_{ss})r^{-(l+2)}e^{im\phi} 
\label{br_pot}
\end{equation}
\begin{equation}
B_\theta^{\rm pot}   =   -   \sum^N_{l=1}\sum^l_{m=-l} 
            B_{lm}\frac{dP_{lm}(\theta)}{d\theta}g_l(r,r_{ss})r^{-(l+2)}    e^{im\phi}
\label{btheta_pot}
\end{equation}
\begin{equation}
B_\phi^{\rm pot}   =  - \sum^N_{l=1}\sum^l_{m=-l} 
                        B_{lm}\frac{P_{lm}(\theta)}{\sin\theta} im g_l(r,r_{ss})r^{-(l+2)} e^{im\phi}.         
\label{bphi_pot}
\end{equation}
The coefficients $B_{lm}$ are determined by the radial component of the surface field.  The functions $ f_l(r,r_{ss})$ and $g_l(r,r_{ss})$  which describe the modification of the field structure by the wind are given by
\begin{equation}
 f_l(r,r_{ss}) = \left[ 
        \frac{l+1+ l(r/r_{ss})^{2l+1}}{l+1+l(1/r_{ss})^{2l+1}}
            \right]
\end{equation}
\begin{equation}
 g_l(r,r_{ss}) =  \left[
       \frac{1 - (r/r_{ss})^{2l+1}}{l+1+l(1/r_{ss})^{2l+1}}
               \right].
\end{equation}
To model a completely closed field, we take the limit $r_{ss} \rightarrow \infty$ and hence $f_l(1) \rightarrow 1$, $g_l(1) \rightarrow 1/(l+1)$.
Only one component of the surface magnetic field is required to determine the unknown  coefficients $B_{lm}$ - typically, the radial component is used. We have adapted a code that was originally developed by \citet{1998ApJ...501..866V} to perform the field extrapolation.  Using this field structure, we determine the location of stable points, the pressure distribution and the prominence locations.

%-------------------- Figure 6 -----------------------
% From plotprom_vels.pro using o/p from allsp3_stable_plus_wind_21May19.pro------------
\begin{figure*}
    \includegraphics[width=3.6cm]{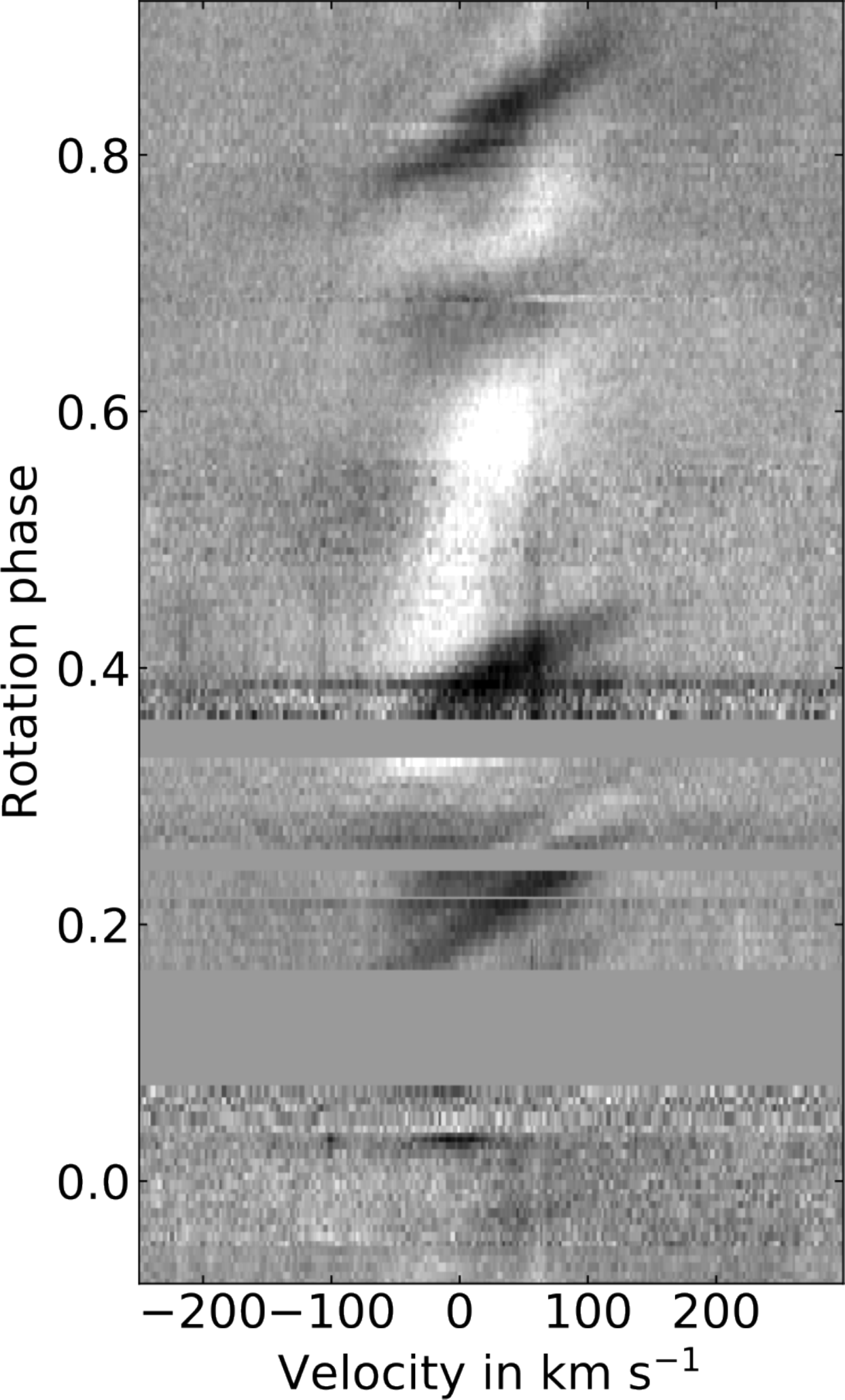}
        \includegraphics[width=3.735cm]{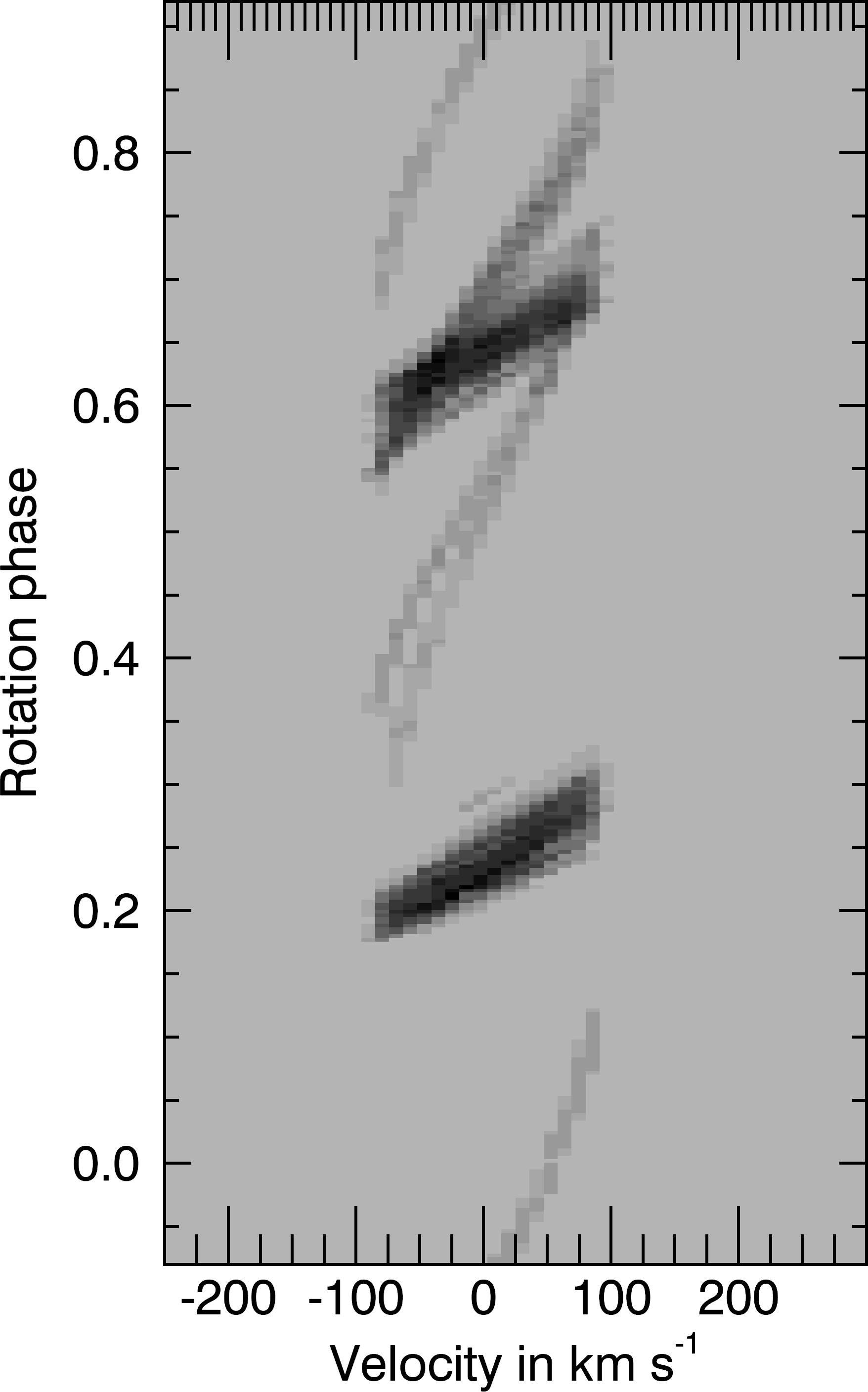}
    \includegraphics[width=3.6cm]{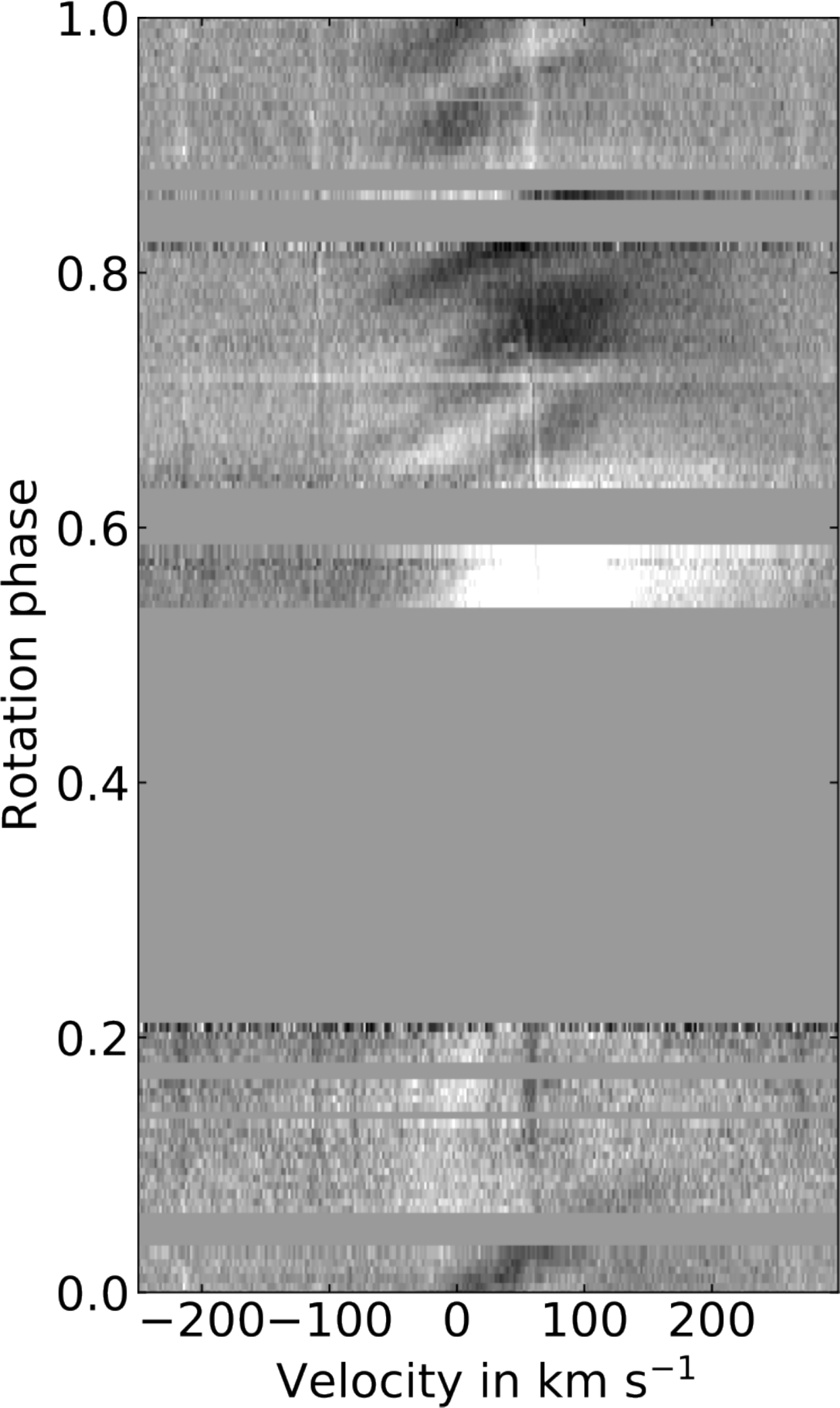}
        \includegraphics[width=3.735cm]{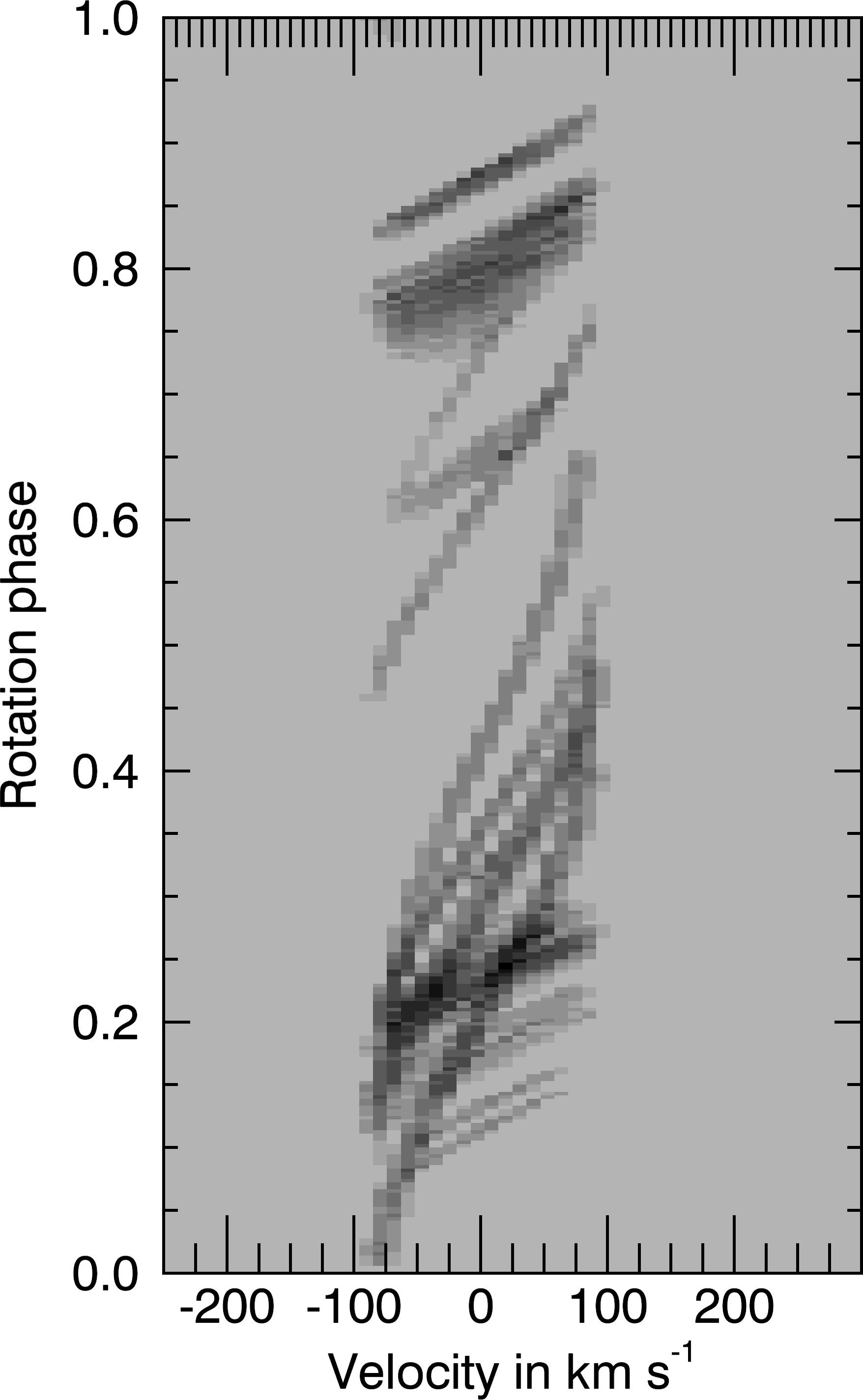}\\
    \includegraphics[width=3.735cm]{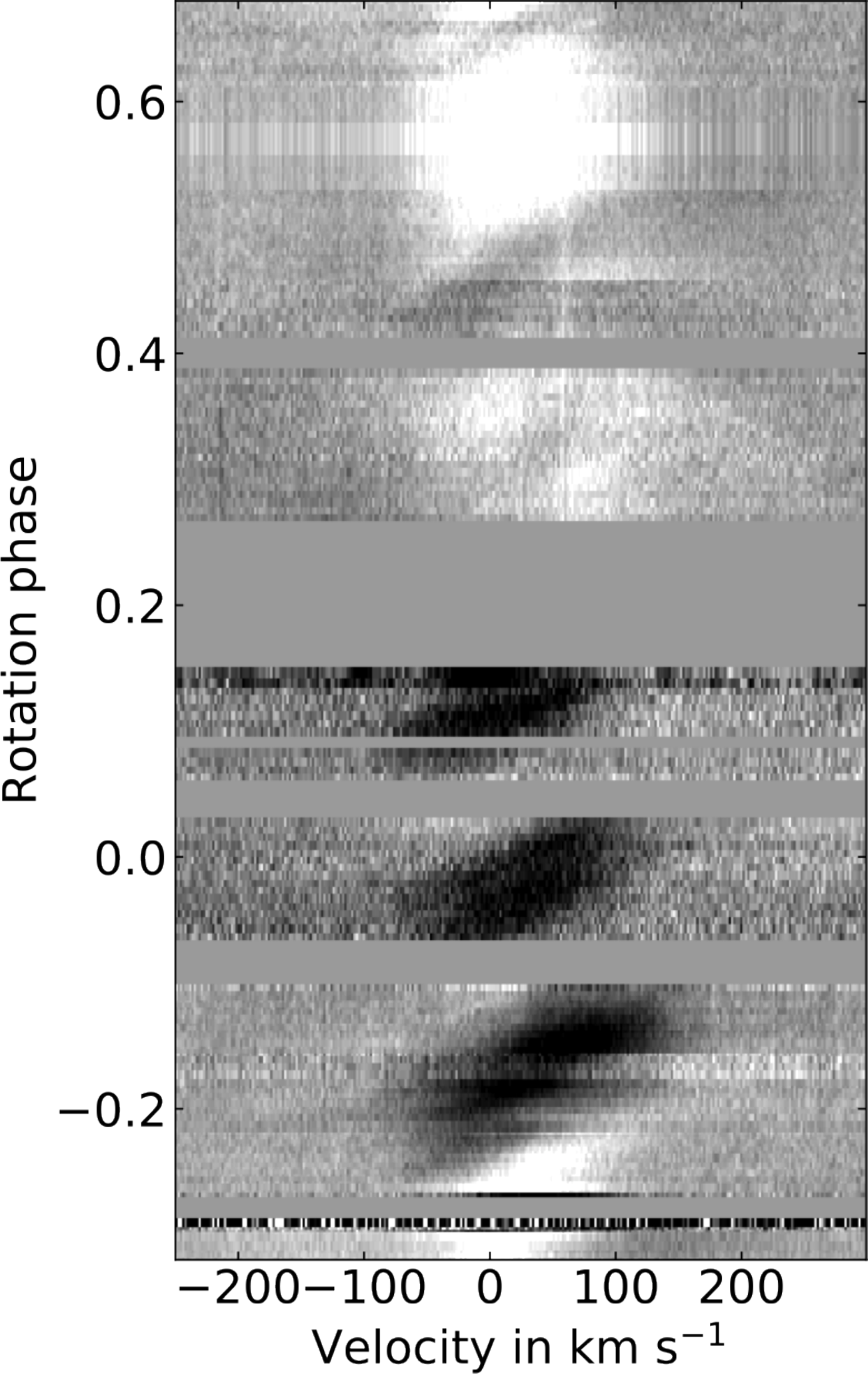}
        \includegraphics[width=3.735cm]{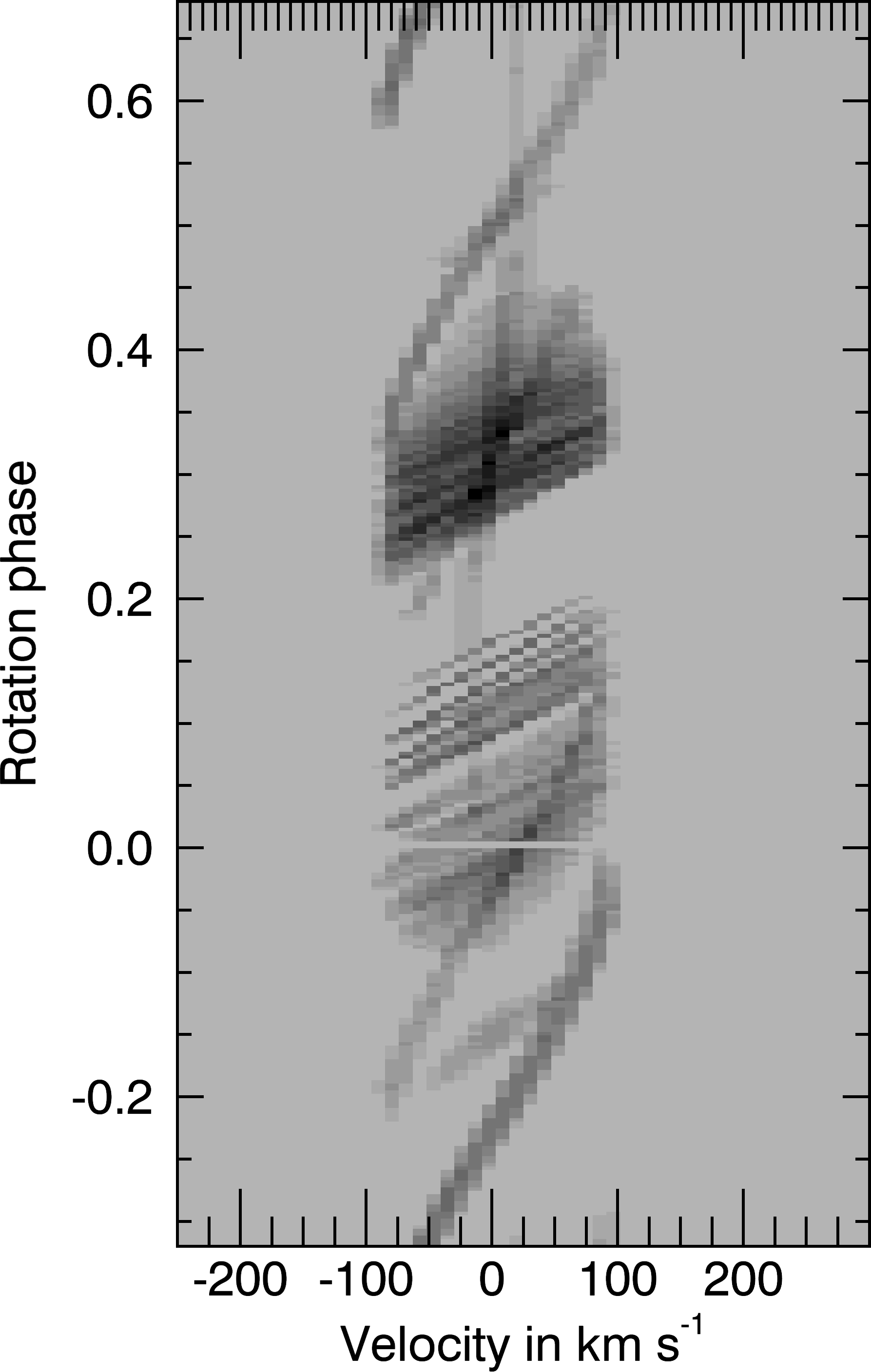}
    \includegraphics[width=3.735cm]{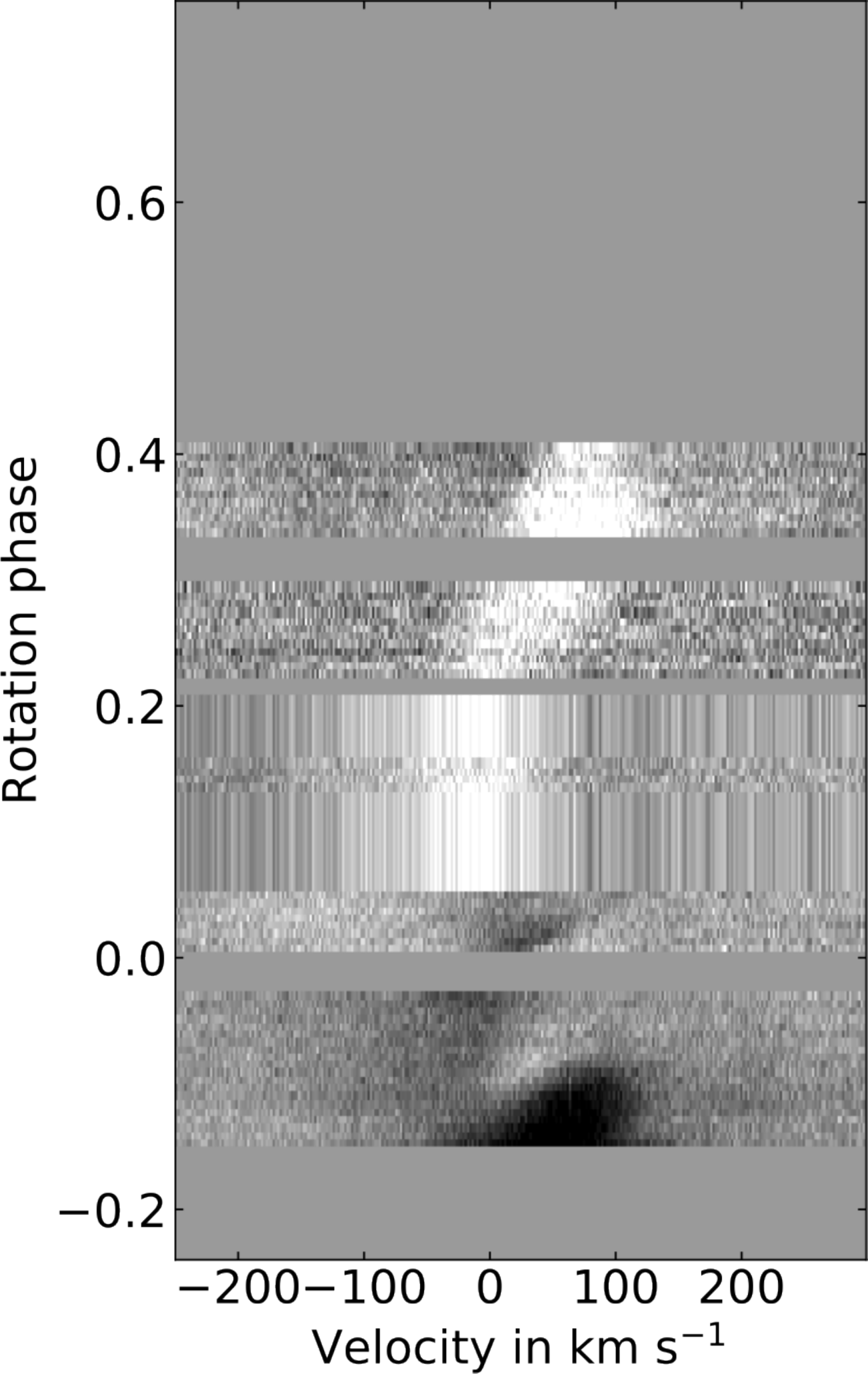}
       \includegraphics[width=3.735cm]{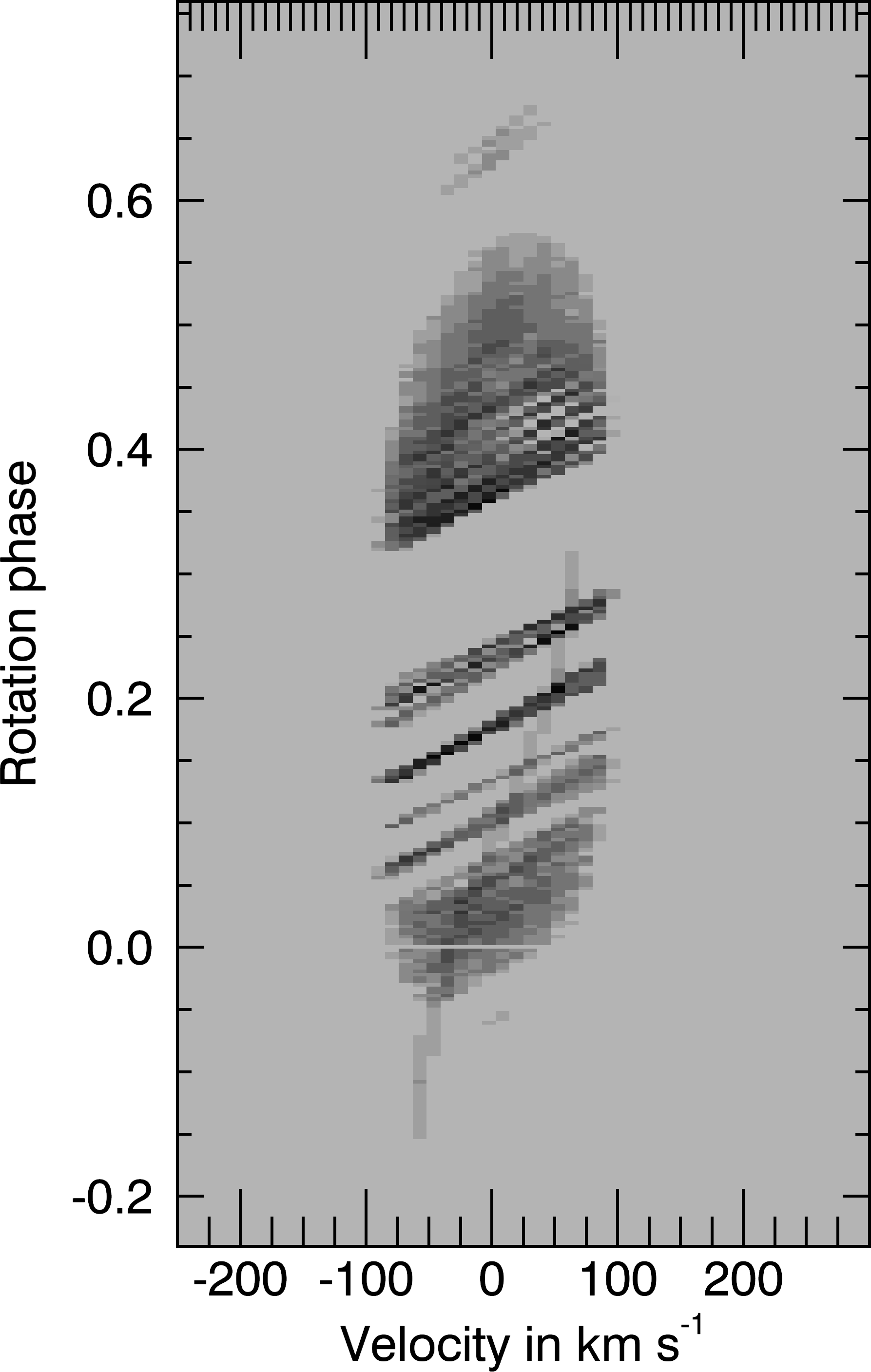} \\
    \includegraphics[width=3.735cm]{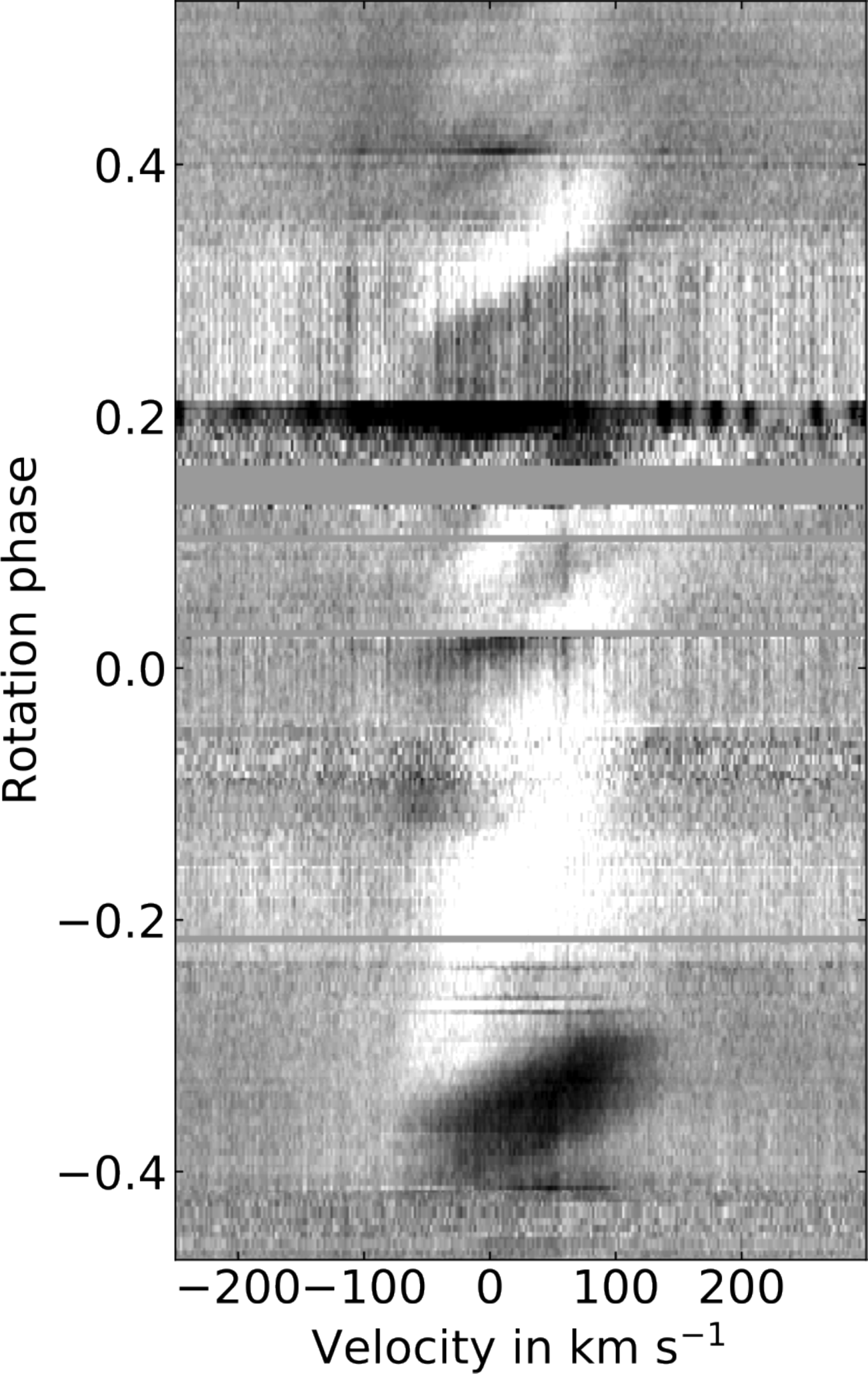}
       \includegraphics[width=3.735cm]{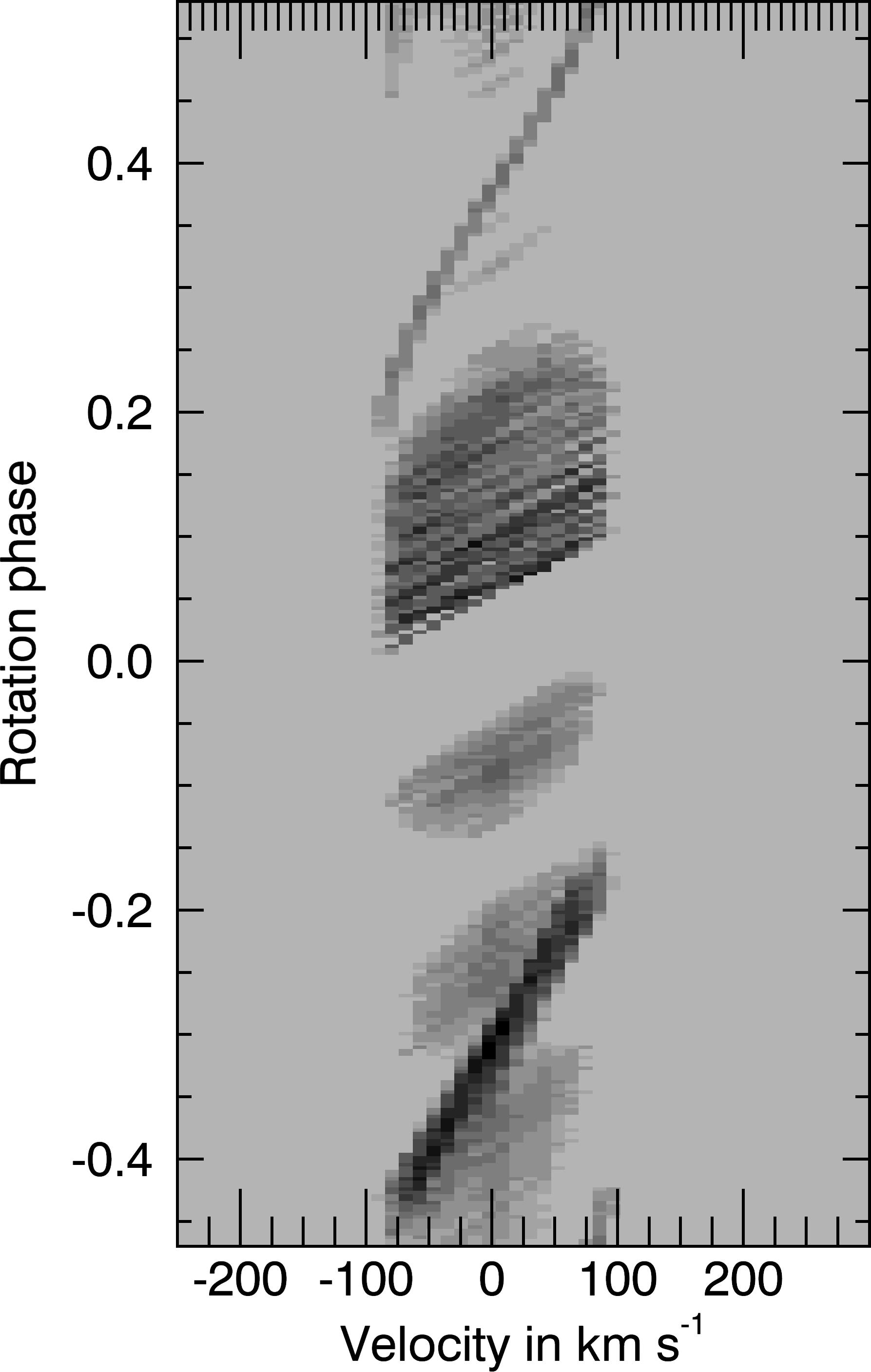} 
    \includegraphics[width=3.6cm]{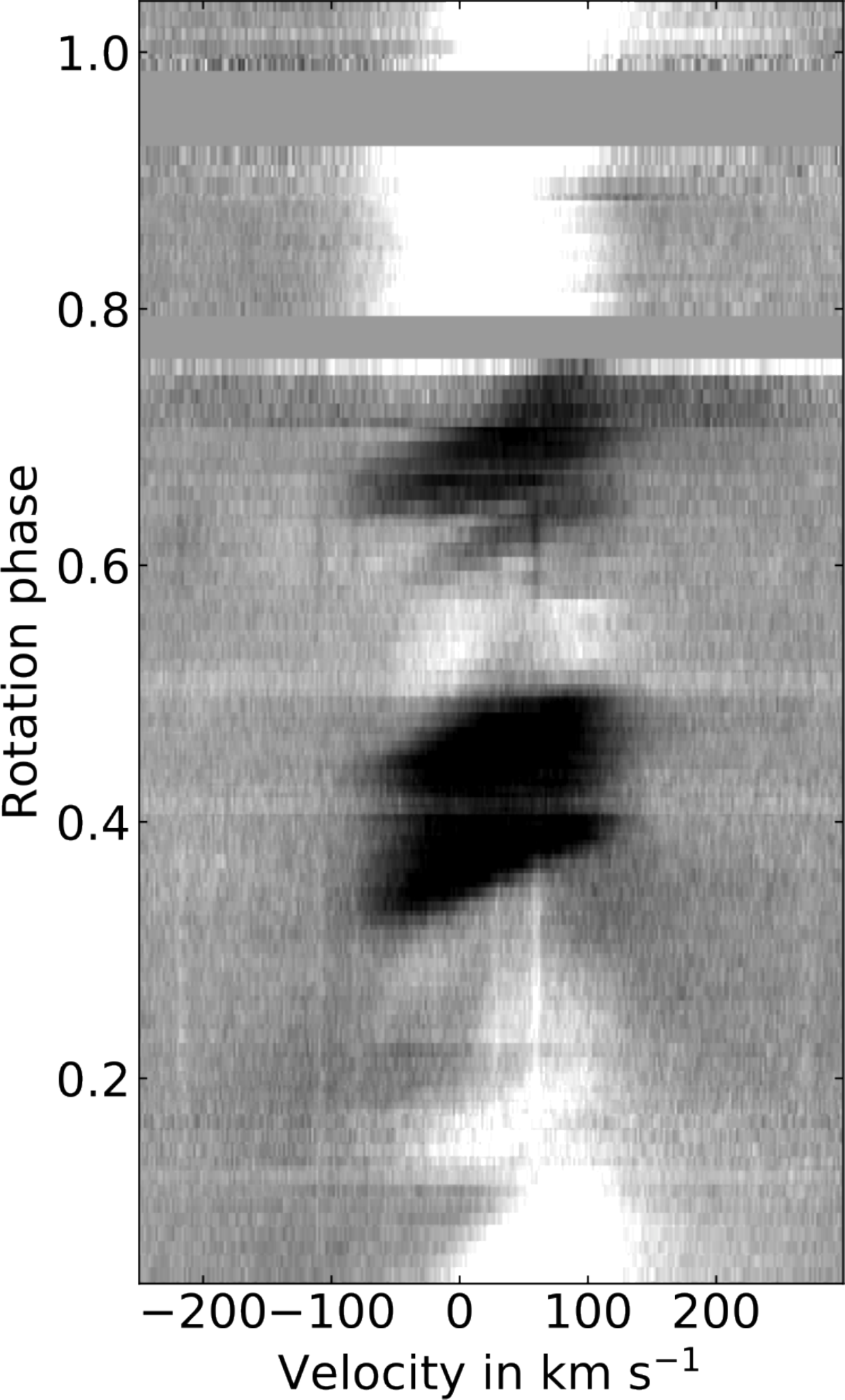}
        \includegraphics[width=3.735cm]{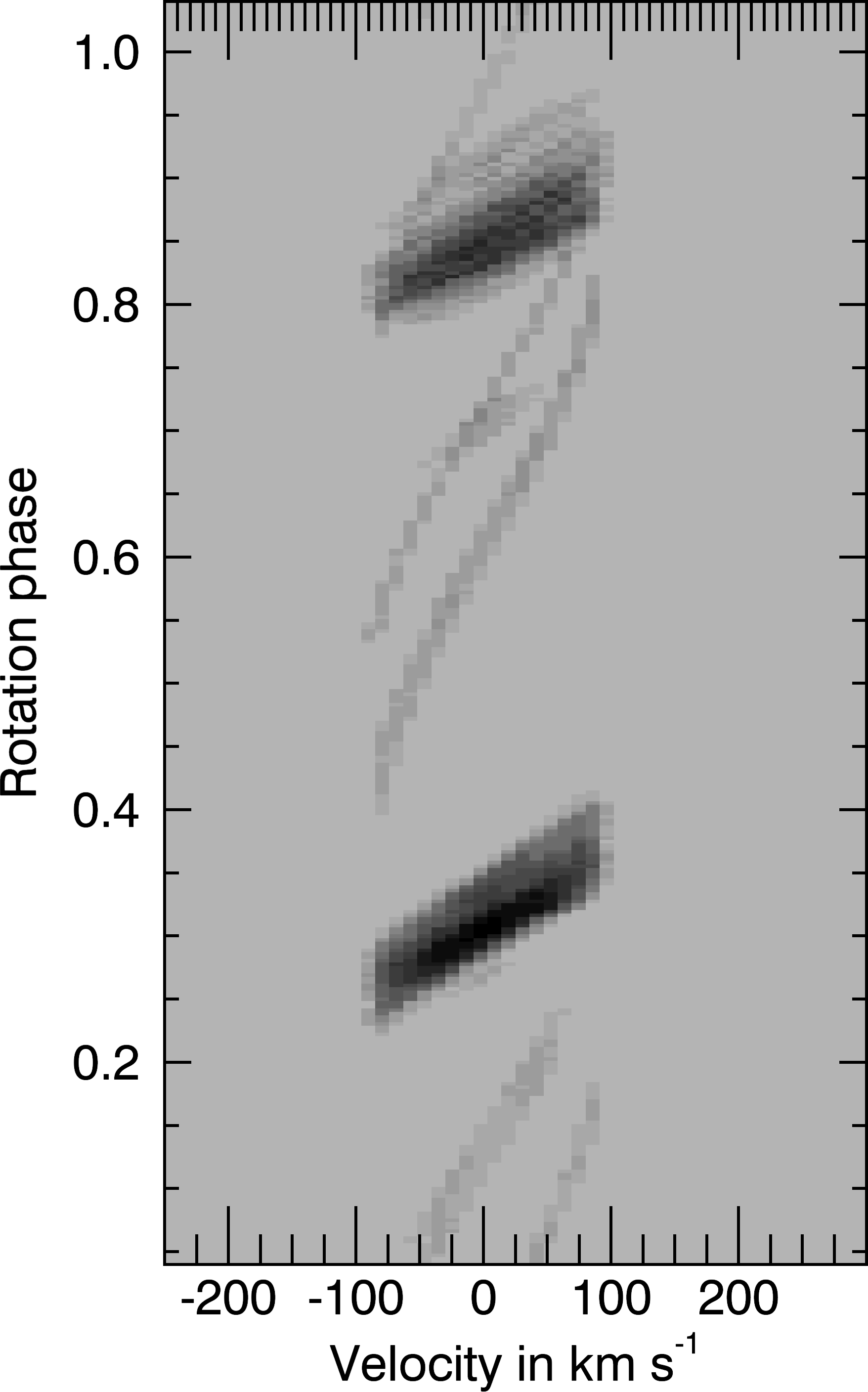}

    \caption{H-$\alpha$ stacked spectra for AB Dor for (top row, left to right) Dec 1995 observed, Dec 1995 synthetic, Jan 1998 observed, Jan 1998 synthetic; (middle row, left to right) Dec 1998 observed, Dec 1998 synthetic, Dec 2000 observed, Dec 2000 synthetic; (bottom row, left to right) Dec 2004 observed, Dec 2004 synthetic, Dec 2007 observed, Dec 2007 synthetic.}
 %       \caption{Maps of the global relative log likelihood $\delta\sum\ln{\mathcal{L}}$ of the observed H-$\alpha$ prominence absorption distribution given the dipole axis locations. The maximum value is shown by a green circle. Also show as a green star is the location of the dipole axis derived from the magnetic maps. }
    \label{fig:all_maps}
\end{figure*}
% End figure ----------------------------------------
%--------------------------------------------------------

The resulting field structure is shown in Fig.~\ref{fig:1996}. The cool prominence material is predominantly supported near the co-rotation radius, at two longitudes separated by around 180 degrees. This is reflected in the morphology of the synthetic H-$\alpha$ transients whose slope and separation matches well the observed features. In contrast to the example of a simple dipole field, where the field line curvature only allows stable equilibria at heights close to and beyond the co-rotation radius, in this case there is also a small amount of cool material supported at lower heights. These take longer to pass through the line profile and so appear in the synthetic H-$\alpha$ spectra as more slowly evolving features with a steeper gradient. We note that the  observed H-$\alpha$ spectra also show bright underlying chromospheric emission features that we do not model.

Fig.~\ref{fig:stars_mprom} shows the resulting variation of the prominence mass with the latitude of the dipole component of the field. The corresponding results for the simple inclined dipole are shown as faint background features for comparison. The prominence masses derived from the magnetograms are consistent with the range of values of $(2-6) \times 10^{14}$kg reported from observations \citep{1990MNRAS.247..415C} but show a spread of two orders of magnitude. This spread is similar to that shown by the simple dipole case, but with a greater scatter due to the variation in the field strength from one epoch to the next.

The lower panel of Fig.~\ref{fig:stars_mprom} also shows the fraction of the total prominence mass that transits the star and so can be detected. As was the case with the simple inclined dipole model, a high dipole inclination (where the dipole axis lies at a low latitude) gives the greatest fraction of mass visible, but the lowest mass supported.

Fig.~\ref{fig:all_maps} shows examples of the observed and synthetic H-$\alpha$ stacked spectra. The synthetic spectra show both slowly-drifting features caused by prominences close to the surface, and rapidly-drifting features due to prominences close to the co-rotation radius. These show similar drift rates to the observed spectra, typically showing absorption at similar rotation phases as is observed. The presence of multiple strands of absorption within each clump is not fully recovered however (typical examples are 1995, and 1996). The H-$\alpha$ stacked spectra are composites, using observations over several nights of observations and so at any one time not all of the absorption features may be present. There are two main features that could mask the presence of some of the absorption in the observed spectra. One is the presence of gaps in the observations (apparent in 1997 and 2000) and the other is the presence of bright emission features (apparent in 1998, 2004 and 2007). We note that the prominence model assumes that all possible prominence support sites are occupied, whereas in practice only a subset will support prominences at any one time. The spectra for 2007 are particularly interesting as two bright emission features are present at the phases where the synthetic spectra show absorption (between 0.8-0.9 and between 0.1 - 0.3) and the two dominant absorption transients appear instead closer together in phase. One possible explanation is that the observations have recorded the aftermath of some energetic event that disturbed the prominences. 

\subsection{A nonpotential field}

While the agreement between the synthetic and observed  H-$\alpha$ spectra is very good, it is worth exploring the impact of our assumption that the magnetic field is potential. We can allow for the presence of non-potential field by writing the total magnetic field as the sum of potential and non-potential components, such that $\underline{B}=\underline{B}^{\rm pot} + \underline{B}^{\rm np}$. Whereas the potential field contribution is completely specified by the choice of source surface $r_{ss}$ and by the surface radial field, there are many possible ways to extrapolate the non-potential field. We choose to select the same form as \citet{2013MNRAS.431..528J} which has a solution in term of spherical harmonics and has a non-potential part that matches the observed surface toroidal field but vanishes at the source surface. Briefly, we make two simplifying assumptions: that the non-potential field lies on spherical shells (ie $B_r^{\rm np} =  0$) and that the electric currents can be derived from a potential $Q$:
\begin{equation}
\nabla \times {\bf B}^{\rm np} = - \nabla Q . 
\end{equation}
As a result, $\nabla^2 Q = 0$, and
\begin{equation}
B_r^{\rm np} =  0 
\end{equation}
\begin{equation}
B_\theta^{\rm np}   =   -   \sum^N_{l=1}\sum^l_{m=-l} 
           l(l+1) C_{lm} \frac{P_{lm}(\theta)}{\sin\theta}imh_l(r,r_{ss})r^{-(l+1)}    e^{im\phi}
\label{btheta_np}
\end{equation}
\begin{equation}
B_\phi^{\rm np}   =   \sum^N_{l=1}\sum^l_{m=-l} 
                l(l+1) C_{lm}\frac{dP_{lm}(\theta)}{d\theta} h_l(r,r_{ss})r^{-(l+1)}  e^{im\phi}   
\label{bphi_np}       
\end{equation}
where
\begin{equation}
 h_l(r,r_{ss}) = \left[ \frac{1-(r/r_{ss})^{2l+1}}{l+(l+1)(1/r_{ss})^{2l+1}} \right]
\end{equation}
and as $r_{ss}\rightarrow \infty$ we recover $h_l(1)\rightarrow 1/l$.

\begin{figure*}
        \includegraphics[width=8.5cm]{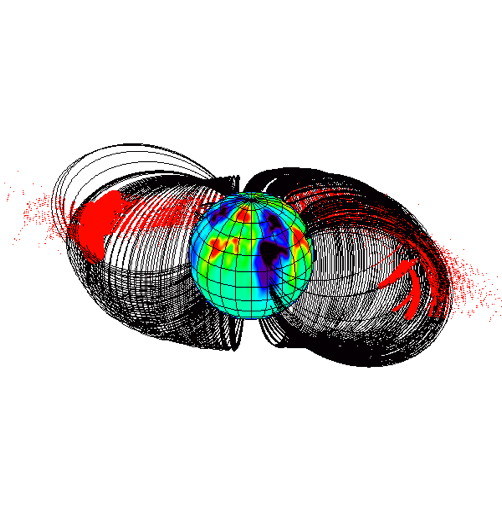}
        \includegraphics[width=8.5cm]{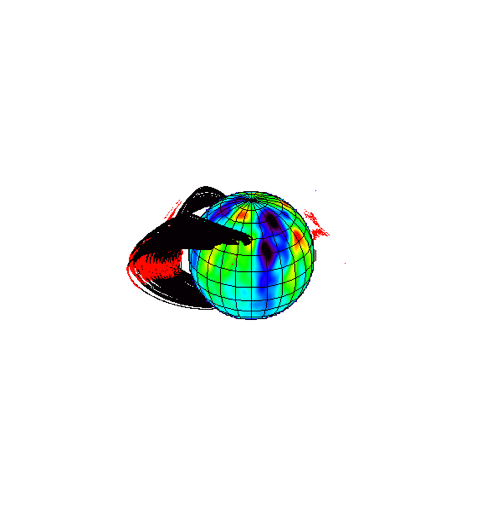}\\
        \includegraphics[width=4.56cm]{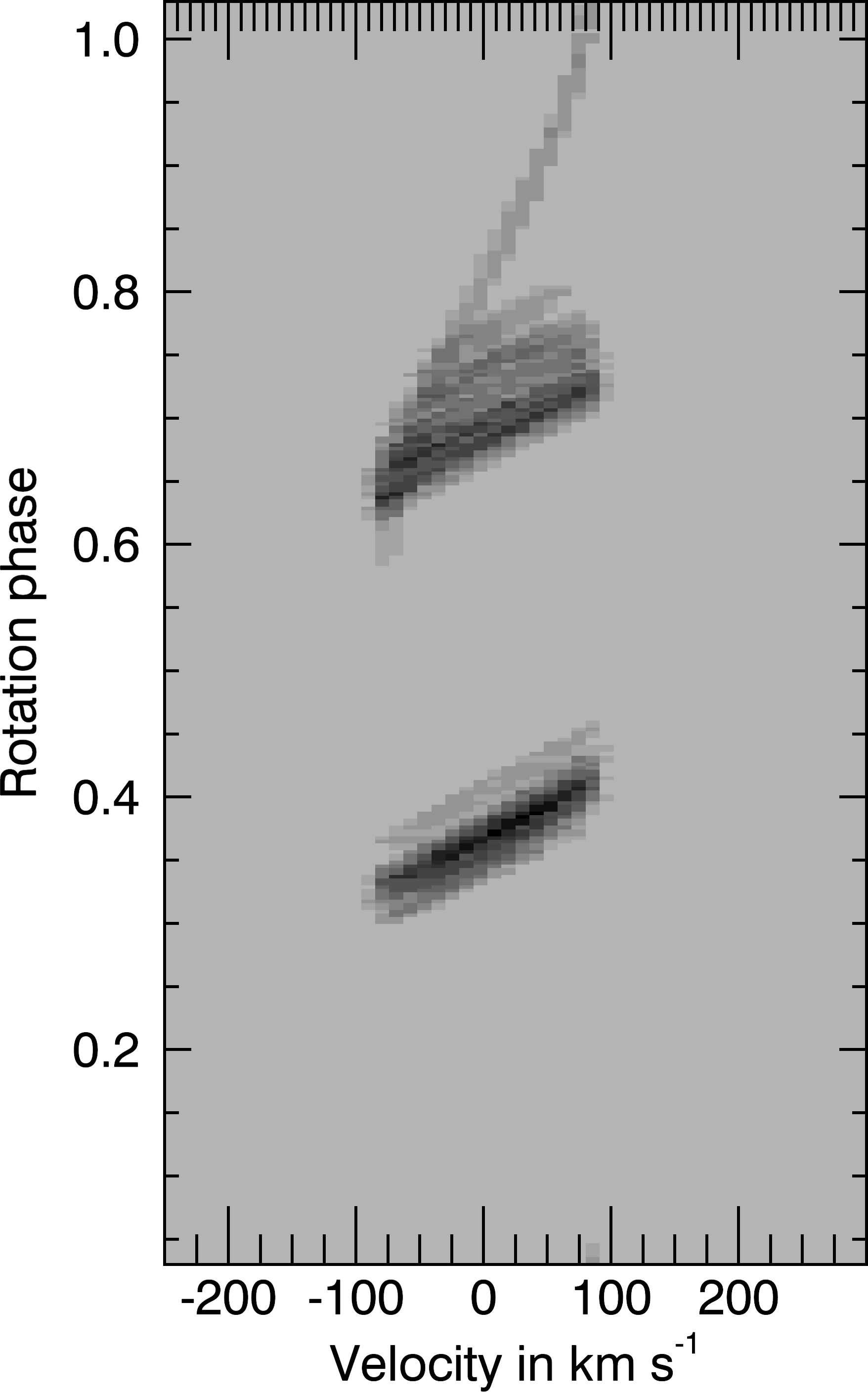}
        \includegraphics[width=4.4cm]{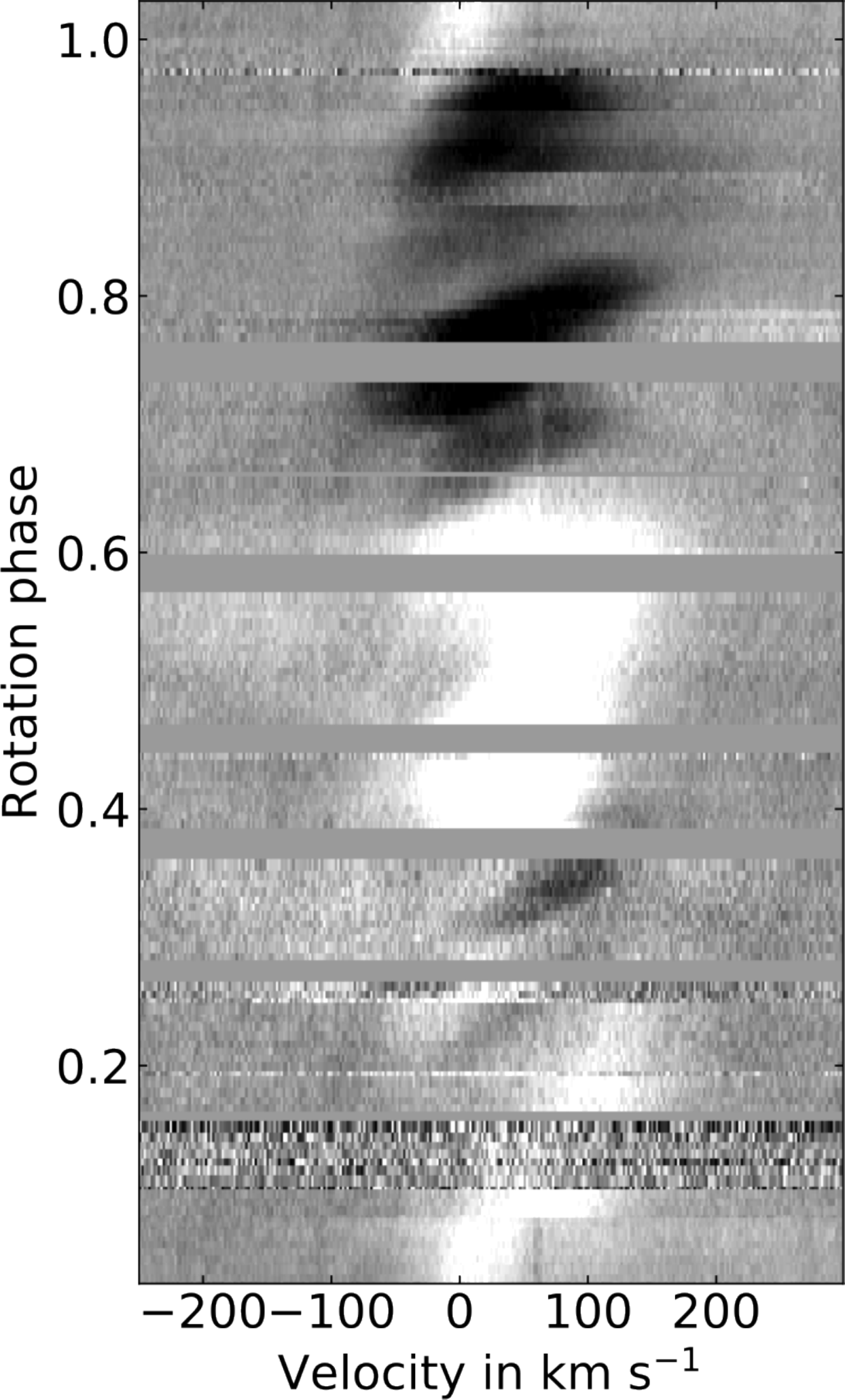}
        \includegraphics[width=4.56cm]{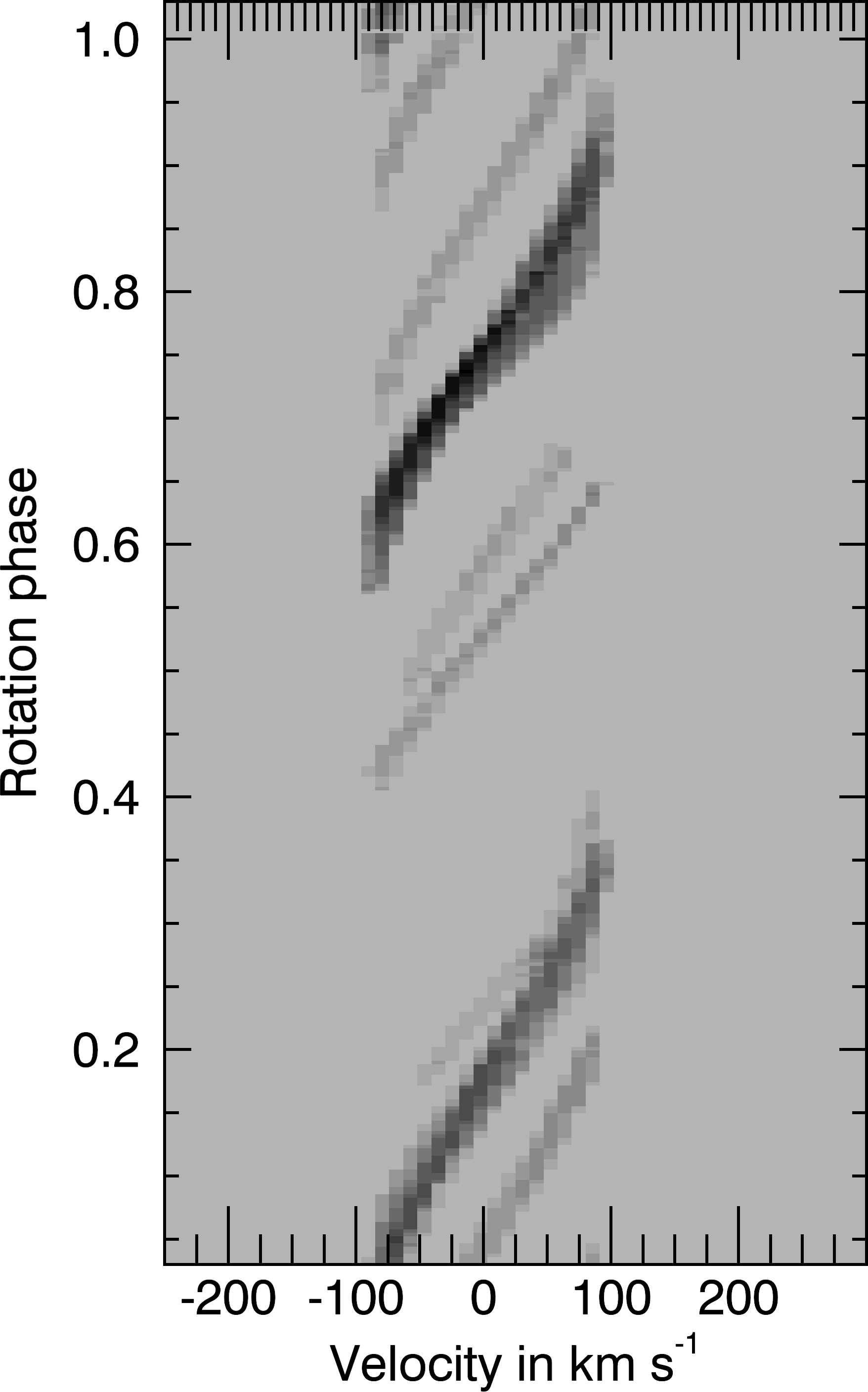} 
    \caption{This figure compares the synthetic H-$\alpha$ spectra produced with potential and non-potential field extrapolations of the surface magnetograms of AB Dor in Dec 2001. The top row shows the large scale field structure for potential (left) and non-potential (right) field extrapolations respectively.  A map of the radial magnetic field is painted on the stellar surface (blue denotes negative field, while red is positive). The cool prominence material is show in red, and the field lines supporting it are drawn in black. The bottom row shows the corresponding synthetic H-$\alpha$ dynamic spectra for a potential field (left) and a non-potential field (right). The observed H-$\alpha$ dynamic spectra are shown in the middle.}
    \label{fig:Collage_pot_nonpot}
\end{figure*}

Fig. \ref{fig:Collage_pot_nonpot} shows the resulting distribution of cool prominence material for both the potential and non-potential field extrapolations and the corresponding H-$\alpha$ spectra for Dec 2001. In the non-potential field case, the cool material is supported at much lower radii and so the absorption transients have a much lower radial acceleration and so take much longer to pass through the line profile. It is clear that the non-potential field extrapolation produces a much poorer match to the observed H-$\alpha$ spectra than does the potential field and we do not consider it further.

 %----------------------Figure 7-------------------
  %---------------------- non-potential field  -------------------
% Figure 7 ---From plotprom_vels.pro and allsp3_stable_plus_wind.pro --
% End figure ----------------------------------------
%--------------------------------------------------------

% SECTION 
%--------------------------------------------------------
\section{Mass and angular momentum loss rates}

%--------------------------------------------------------
% TABLE 1 
%--------------------------------------------------------
%--------------------------------------------------------
\begin{table*}
	\begin{tabular}{lcccc|ccc|r} % 8 columns, alignment for each
		\hline
Year &  $<B^2>$ & i & m$_{\rm tot}$  & m$_{\rm vis}$ & $\dot{m}_p$ & $\dot{m}_w$ & $\dot{j}_p$ & $\dot{j}_w$ \\
     & [$10^4$ G$^2$] & [$^\circ$] & [10$^{14}$ kg] & [10$^{14}$ kg]   & [10$^{-14}$M$_\odot$/yr] & [10$^{-14}$ M$_\odot$/yr] & [10$^{32}$ erg] & [10$^{32}$ erg]\\
		\hline
		1995 & 4.42 & 15 & 0.28  & 0.10   & 1.44  & 2.77 & 1.26 & 9.90 \\
		1996 & 2.25 & 30 & 0.36  & 0.07   & 2.23  & 3.53 & 1.74 & 3.64 \\
		1997 & 3.65 & -35 & 0.87  & 0.18   & 0.38  & 14.75 & 0.24 & 10.73 \\
		1998 & 3.70 & 45 & 10.01 & 1.83   & 21.44 & 63.45 & 17.49  & 36.93 \\
		1999 & 1.47 & 26 & 0.20  & 0.02   & 0.60  & 1.99 & 0.52 & 4.60 \\
        2000 & 3.32 & 55 & 1.40  & 0.03   & 6.84  & 18.99 & 5.93 & 9.89\\
        2001 & 6.87 & 40 & 0.79  & 0.09   & 4.42  & 17.08 & 4.24 & 14.20 \\
        2002 & 4.75 & 12 & 0.95  & 0.31   & 2.65  & 15.14 & 2.55 & 54.00\\
        2003 & 2.10 & 0 & 0.09  & 0.04   & 0.35  & 0.78 & 0.35 & 5.69 \\
        2004 & 5.69 & 57 & 14.55 & 0.66   & 23.91 & 58.22 & 20.20 & 27.57 \\
        2007 & 4.57 & -11 & 29.09 & 1.34   & 2.28 & 33.71 & 2.26 & 164.03 \\  
		\hline
	\end{tabular}
    \caption{Derived prominence properties based on the Zeeman-Doppler maps for each year. Columns show the year, the mean squared flux density, the inclination of the dipole axis, the total mass supported in prominences, the prominence mass that is visible, the mass loss rate in ejected prominences, the mass loss rate in the wind alone (excluding prominences) and the associated angular momentum loss rates in the prominences and the wind.}
    \label{table:data}
\end{table*}
%--------------------------------------------------------
%--------------------------------------------------------
% SUBSECTION 
%--------------------------------------------------------
\subsection{Prominence mass loss rates}
%--------------------------------------------------------

These estimates of prominence masses suggest that prominence ejection may provide a significant contribution to the total mass loss from the star. Observing a prominence ejection is however a low-probability event.  \citet{DunstoneThesis2008} present one such observation, but comment that since the prominences can only be detected in absorption when the pass in front of the star (which may take only an hour or so) then if they have lifetimes of order a few days, then the window within which they can be observed comprises only a few percent of their lifetimes. One prominence ejection observed among 70 prominences detected is therefore consistent. 

\citet{2019MNRAS.482.2853J} describe how, in stars with sufficiently hot coronae,  the upflow that forms slingshot prominences may be supersonic by the time it reaches the prominence formation site. In this case, the surface is unable to respond to the formation of a prominence and so continues to supply a mass upflow, increasing the prominence density until it becomes too great for magnetic support. At this point, any prominence material above the co-rotation radius will be centrifugally expelled. This limit-cycle behaviour essentially provides an intermittent wind loss from the star from the prominence-bearing loops. The rate of mass loss is determined by the rate at which the surface can supply mass. For a thermal (Parker-type) wind this is determined by the temperature. Close to the stellar surface the velocity of a thermal wind has the asymptotic form \citep{1958ApJ...128..677P,1999isw..book.....L}
\begin{equation}
u(r_\star)=c_s(r_{ss}/r_\star)^2e^{3/2}e^{-2(r_{ss}/r_\star)},
\label{parker}
\end{equation}
where the sound speed is given by $c_s^2 = kT/m$ for a temperature $T$ and mean particle mass $m$. The sonic radius $r_{ss}$  where the wind speed reaches the sound speed (and so  $u=c_s$) is given by
\begin{equation}
r_{ss} = \left( \frac{GM_\star}{2c_s^2} \right).
\label{sonic_radius}
\end{equation}
The mass loss rate from each prominence is then
\begin{equation}
\dot{m}_p = A_\star\rho_\star u_\star
\label{Mdot}
\end{equation}
where $\rho_\star$ is the mass density at the stellar surface and $A_\star$ is the footpoint area of the prominence-bearing loop.

There are two free parameters in this expression - the temperature and the base density. The base density is determined from $\rho = p/c^2$ where the base plasma pressure $p$ is given by $p=\kappa_p B^2$. The base density and hence the mass loss rate therefore scale linearly with $\kappa_p$. The dominant effect of the temperature on the mass loss rate is through its influence on the wind speed. Increasing the temperature increases the wind speed more than it reduces the density, with the result that the mass loss rate increases with temperature. 

\begin{figure}
\begin{centering}
    \includegraphics[width=\columnwidth]{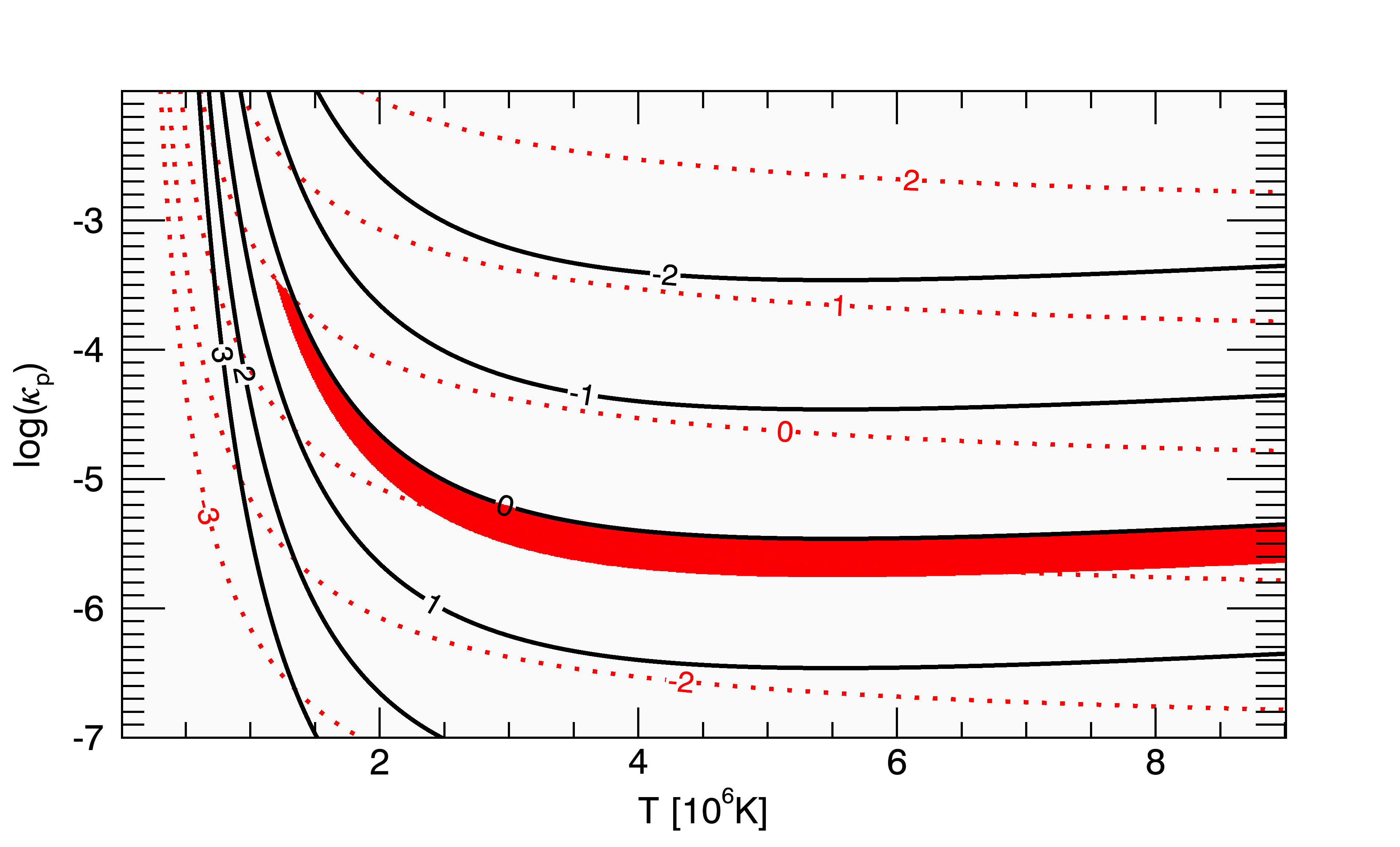}
    \caption{Contours of the logarithms of the prominence lifetimes (black, solid) and plasma $\beta$ at the co-rotation radius (red, dotted) for a range of value of the free parameters $\kappa_p$ and loop temperature $T$. Within the red shaded region, the prominence lifetimes lie in the range $1 < \tau[\rm days] < 10$ and the plasma $\beta < 1$. }
   \label{fig:contours}
\end{centering}
\end{figure}

We can use the prominence observations to provide reasonable values for these two parameters. If $\kappa_p$ is too large, the plasma pressure will exceed the magnetic pressure at some height and the coronal gas will not be confined. We assume therefore that at the co-rotation radius, where we observe that prominences are confined, the plasma pressure must be less than the magnetic pressure ($\beta<1$). Since the prominence mass derived from (\ref{eq:rho_max}) is independent of either $\kappa_p$ or the temperature, the prominence lifetimes $\tau = m_p/\dot{m}_p$ depend on $(\kappa_p, T)$ only through the influence of these parameters on $\dot{m}_p$. We can use the observed prominence lifetimes of $\tau \simeq 1$ day \citep{collier1989II} to constrain $\kappa_p$ and $T$. Fig. \ref{fig:contours} shows contours of the logarithm of the plasma beta at the co-rotation radius and also the lifetime $\tau$ as a function of the two free parameters. Only areas of this parameter space with $\beta<1$ and $\tau \simeq 1$day are acceptable. These are shown shaded. We note that for temperatures above around $3 \times 10^6$K these constraints are insensitive to the temperature. The restriction to prominence lifetimes of around 1 day provides however a narrow range of values of $\kappa_p$. We select a value of $\kappa_p=10^{-5.5}$ such that even at the lowest temperature (T=$2.5\times 10^6$K) at which the upflow is still supersonic at the co-rotation radius, the prominences still have a lifetime of 1 day. We select the temperature of $8.57 \times 10^6$K derived from the overall stellar X-ray flux \citet{2015A&A...578A.129J}

%As commented in  \citep{2019MNRAS.482.2853J}, it is unlikely to be lower than the temperature of the quiescent solar wind ($3.8\times 10^6$K and $1.8\times 10^6$K for the fast and slow solar wind respectively \citep{2015A&A...577A..27J}). It is also unlikely to be significantly higher than the value of $8.57 \times 10^6$K derived from the overall stellar X-ray flux \citet{2015A&A...578A.129J}. 

%--------------------- Figure 9 -------------------
% Figure 9 ----From (plot)prom_mass_time.pro and plotprom_stats.pro--
% End figure ----------------------------------------
%--------------------------------------------------------

%------------------ Figure 10 ----------------------
% Figure 10 ----From (plot)prom_mass_time.pro and plotprom_stats.pro--
\begin{figure}
\begin{centering}
    \includegraphics[width=\columnwidth]{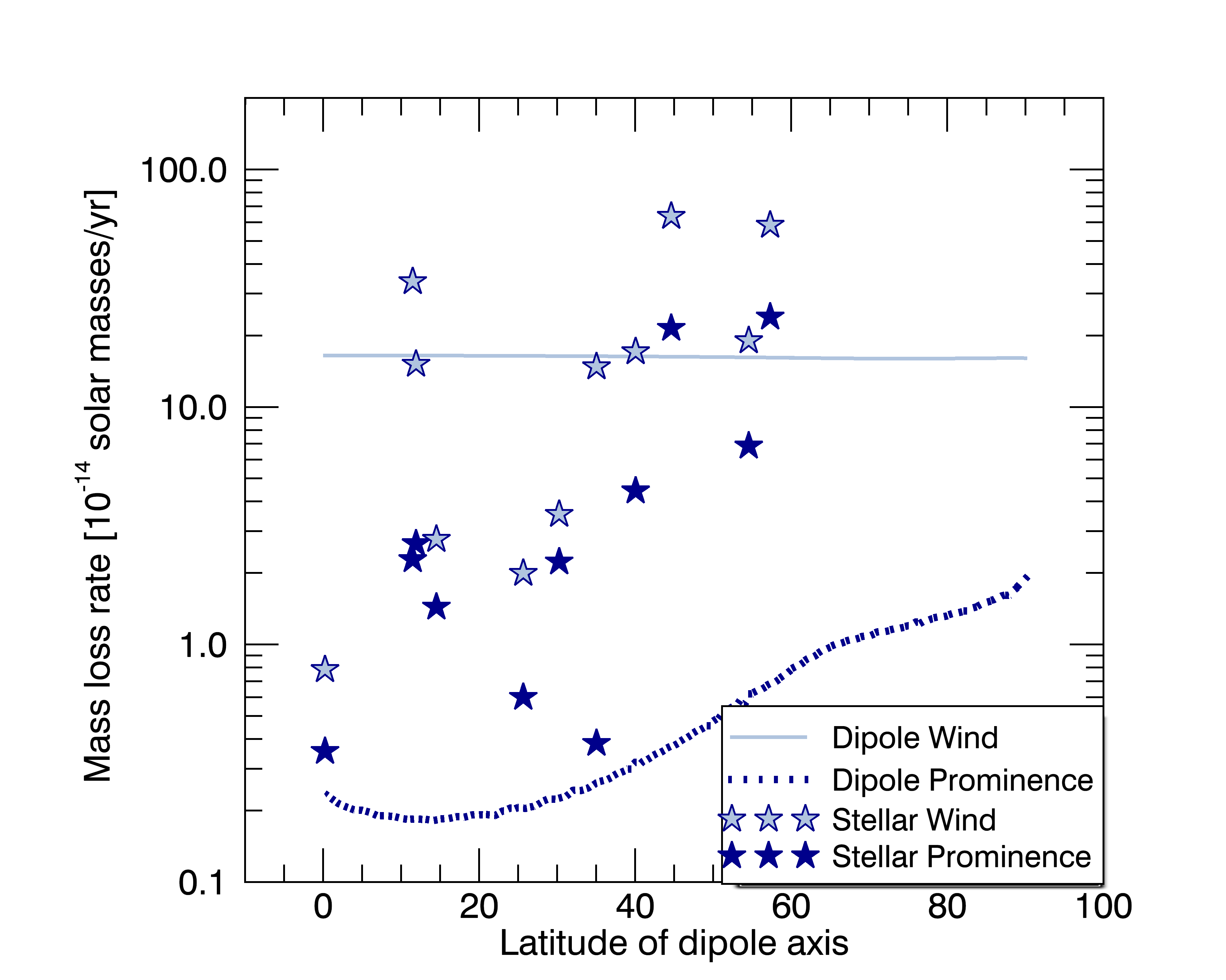}
    \caption{The contributions to the mass loss rate from various sources. Stars denote values based on observations of AB Dor over 10 observing seasons, while lines denote values based on a series of inclined dipoles with a polar field strength of 40G. For each observing season for AB Dor, two sources are shown. Values for AB Dor's prominence system are shown as dark blue stars, while values for AB Dor's wind are shown as faint blue stars.  The faint blue solid line shows wind losses from a purely dipolar field, while the dark blue dotted line shows the corresponding losses in prominences.}
    \label{fig:stars_mdot}
\end{centering}
\end{figure}
% End figure ----------------------------------------

%--------------------- Figure 11 -------------------
% Figure 11 ----From (plot)prom_mass_time.pro and plotprom_stats.pro--
\begin{figure}
\begin{centering}
    \includegraphics[width=\columnwidth]{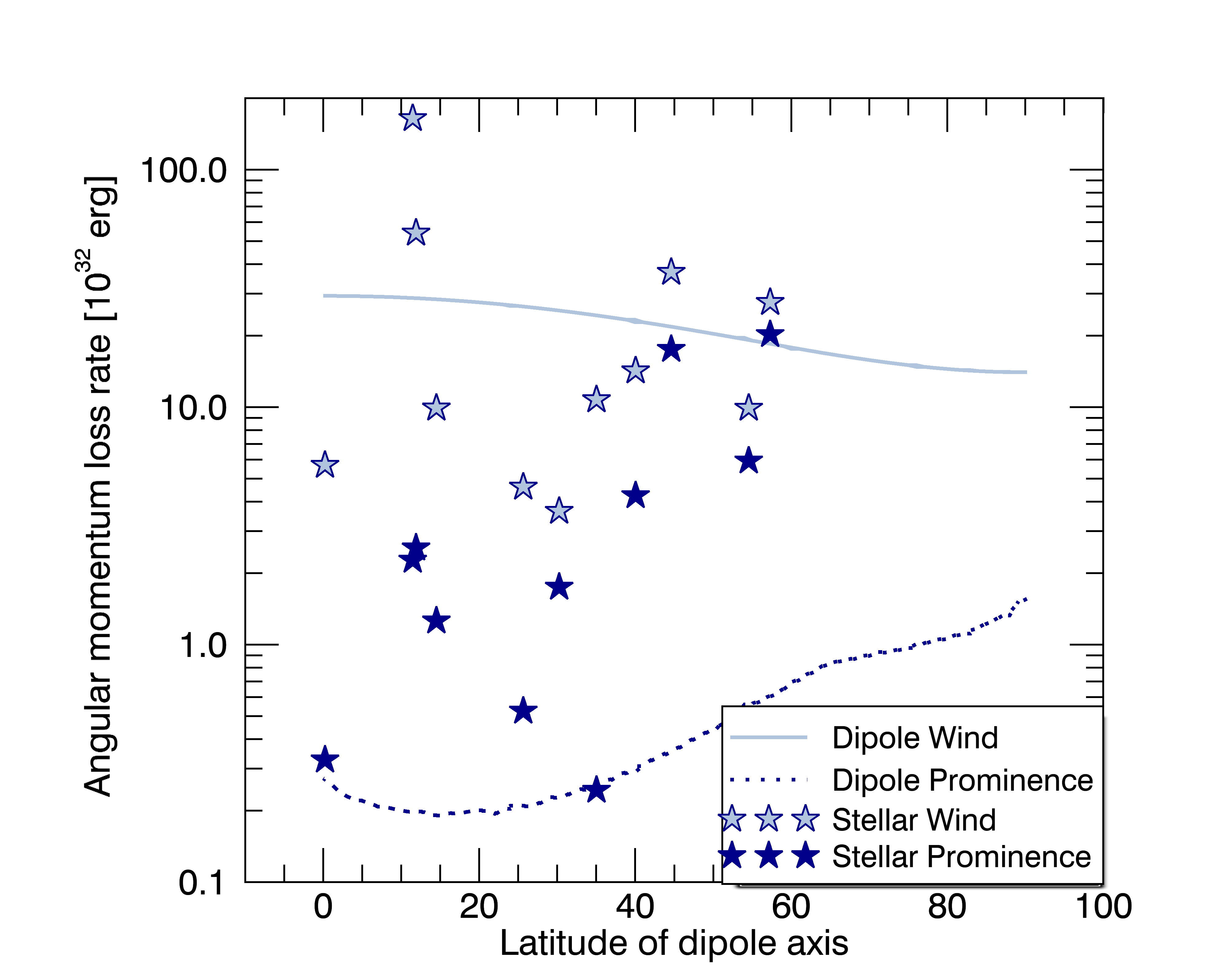}
    \caption{The contributions to the angular momentum loss rate from various sources. Stars denote values based on observations of AB Dor over 10 observing seasons, while lines denote values based on a series of inclined dipoles with a polar field strength of 40G. For each observing season for AB Dor, two sources are shown. Values for AB Dor's prominence system are shown as dark blue stars, while values for AB Dor's wind are shown as faint blue stars.  The faint blue solid line shows wind losses from a purely dipolar field, while the dark blue dotted line shows the corresponding losses in prominences.}
    \label{fig:stars_jdot}
\end{centering}
\end{figure}
% End figure ----------------------------------------

Fig. \ref{fig:stars_mdot} shows the resulting mass loss rates obtained by summing over all the prominences (see also Table~\ref{table:data}). Since both the upflow temperature and $\kappa_p$ are kept constant, the year-to-year variation is due primarily to variations in the field strength at the stellar surface. The values for AB Dor (shown as stars) can also be compared with the values for the simple dipole (shown as a bold dotted line). In both cases, a dipole axis that is at the greatest latitude and hence most closely aligned with the rotation axis, provides the greatest prominence support and hence mass loss. In the case of AB Dor, year-by-year variations in the field strength also provide more scatter than in the simple inclined dipole case.

\subsection{Wind mass loss rates}

Fig. \ref{fig:stars_mdot} also shows the mass loss rates carried by the stellar wind flowing along the open field lines. To determine this, we use the method of \citet{2017MNRAS.465L..25J} that is based on the {\it WSA} method of modelling the solar wind \citep{wang1990,arge2000}. The expansion of the magnetic field with height above the stellar surface determines the wind speed through an empirical relation. Thus, for any field line (labelled $i$) the expansion factor $f_i$ is given by:
\begin{equation}
f_i = \frac{r_\odot^2}{r_{\rm{s}}^2} \frac{B_i(r_\odot)}{B_i(r_{\rm{s}})}
\end{equation}
and the velocity of the wind along that field line at the source surface is given by \citep{wang1990,arge2000}
\begin{equation}
u_i [{\rm kms}^{-1}] = 267.5 + \frac{410.0}{f_i^{2/5} }.
\end{equation}
From this we can determine the mass loss rate for a 1D isothermal wind solution along each field line. The requirement that the wind is trans-sonic and reaches the velocity $u_i$ then determines the field line temperature. The base density follows from the relation $p_0=\kappa_w B_0^2$ at the base of the field line where we set the free parameter $\kappa_w$ to a value that produces the variation in the solar mass loss rate through its cycle \citep{cranmer_review_2008}.  Conservation of mass and magnetic flux requires that $\rho u /B$ is constant along each flux tube, providing the mass loss rate through a spherical surface $S$
\begin{equation}
\dot{M}=\oint_{S} \rho_i u_i dS_i
\end{equation}
 where $\rho_i$ is the density at this surface and $dS_i$ is the cross-sectional area of the flux tube.

As can be seen from Fig. \ref{fig:stars_mdot} the mass loss rate carried by the wind is greater than that carried by the prominences. For a simple dipole field, the mass loss rate is insensitive to the dipole inclination, whereas for AB Dor the trend for higher field strength to also have higher dipole latitude gives a rise in mass loss rate with increasing dipole latitude. We note that the scatter in mass loss rates in both the wind and prominences is similar for the stellar fields.

\subsection{Angular momentum loss rates}

In order to calculate the angular momentum loss rates for the prominences and wind, we assume that each prominence removes only the specific angular momentum $L=\Omega_\star \varpi^2_p$ it carries at its formation site (at cylindrical radius $\varpi_p$). This neglects the torques exchanged during the ejection process and so provides a lower limit to the angular momentum loss in the prominences. The wind, by comparison, carries away angular momentum from an {\it effective} radius which is the Alfv\'en radius (the radius where $u(r) = B(r)/\sqrt{\mu \rho(r)}$). We estimate the total angular momentum loss rate in the wind by integrating over the Alfv\'en surface ($S_A$)
\begin{equation}
\dot{J} = \oint_{S_A} \rho (\underline{u}\cdot\underline{n}) \Omega_\star \varpi^2 dS_A
\end{equation}
where $\underline{n}$ is the outward normal. We note that this neglects the small term due to non-axisymmetry described in \citet{mestel_book_99}. 
The angular momentum loss rates for the prominences and the wind are shown in Fig. \ref{fig:stars_jdot}. For the dipole magnetic field, shown as dotted lines, increasing the latitude of the dipole axis leads to an increase in the number of prominence support sites and hence overall prominence mass and angular momentum loss. It leads to an increase in the latitude (and hence a reduction in the lever arm) of the wind-bearing field lines, however, and so a decrease in the angular momentum losses in the wind.  

For the AB Dor field geometries, however, (shown as stars) the angular momentum losses in both the prominences and the wind show the same trend of increase with dipole latitude as the mass loss rates. The amplitude of variation in the values for the wind is less than for the prominences, however. This is because the variations in $B^2$ that drive the variations in the base density of the wind (and hence $\dot{M}$) are suppressed to some degree by the variations in the Alfv\'en radius. When $B^2$ increases, the base density increases, but the Alfv\'en radius decreases. This behaviour affects the angular momentum loss rates in the wind, but not in the prominences.

% SECTION
\section{Discussion}

The observed H-$\alpha$ absorption transients of cool stars typically show drift rates that place the absorbing material at (or beyond) the co-rotation radius. These clumps of absorbing material are estimated to have masses, in the case of AB Dor, of $2-6 \times 10^{14}$kg, some three orders of magnitude more massive than large solar quiescent prominences  \citep{1990MNRAS.247..415C}. We find that both the location and masses of such ``slingshot prominences'' can be reproduced by a potential field extrapolation of the surface magnetic field that is recovered by Zeeman-Doppler imaging. This type of extrapolation of the coronal magnetic field uses only the radial component of the surface magnetic field. It includes  azimuthal and meridional components consistent with the assumption that the field is potential, but does not include the additional non-potential part of the field often present in the magnetograms of active stars. This non-potential component is carried mainly in the azimuthal field \citep{2015MNRAS.453.4301S}. 

The role of this non-potential field in the structure and dynamics of the corona is not known. \citet{2013MNRAS.431..528J} initiated an MHD wind model for two low mass stars (CE Boo and GJ 149) with this field included, but demonstrated that the wind solution relaxed quickly back to a potential field close to the surface.\footnote {Our static model does not include the azimuthal field component produced by the stresses induced by the stellar wind.} The non-potential field observed in the ZDI field maps appeared to have little effect on the star's wind, suggesting that the strong azimuthal fields detected at the surface do not extend significantly into the corona. We have used the same type of non-potential field extrapolation to predict the distribution of prominence mass and hence the appearance of the stacked H-$\alpha$ profiles that would result. Including this strong additional azimuthal field provides extra support for cool mass in the corona and leads to prominence support sites at low heights. These dominantly azimuthal field lines however have a large radius of curvature and cannot support prominences at the height of the co-rotation radius. The resulting absorption transients have a drift rate that is clearly unlike what is observed.

Both the MHD wind models and the prominence models  suggest that the strong non-potential fields detected at the surface of AB Dor and other rapidly-rotating cool stars do not extend out to the very large heights at which prominences are detected. Their associated currents must be confined closer to the stellar surface, where they can power the strong and frequent flares of these stars.

The location of the prominences at heights of several stellar radii suggests that the dipole component of the field (which decays most slowly with height) will dominate their support. We find that the inclination of the dipole axis to the rotation axis has a significant influence on the distribution of prominences. Magnetic fields whose dipole axes are fully aligned can support 100 times more mass that those that are highly inclined. In stars with magnetic cycles similar to the Sun's, where the field reverses over the cycle, we would therefore predict that the mass supported in prominences could vary significantly over the cycle. Since the mass also varies as $B^2$, the decay in field strength with rotation rate that takes place as stars ages and spin down will also reduce the mass that can be supported. For very young stars, such as LQ Lup, where the dipole field component can be as strong as a few hundred G (and up to a kG) prominence masses may be up to 1000 times greater than detected in AB Dor \citep{2000MNRAS.316..699D}. By considering a wind evolution model, \citet{2019MNRAS.485.1448V} examined the evolutionary periods over which prominences could be supported in solar-like stars. The peak in prominence mass loss rate is reached when the star reaches the Zero Age main Sequence, which for a solar mass star is around 40Myrs. If such a star is initially rotating rapidly, prominence ejection may continue until an age of 800Myrs.

We show that the distribution of prominence mass is such that much of it may not pass in front of the star and so would not be seen in absorption transients. While almost 50$\%$ of the mass may be detected in  systems where the magnetic and rotational axes are highly inclined, this fraction falls to as little as a few percent for highly aligned cases. Estimates of the overall mass supported in prominences that are based on sparse observations of only a few prominences may therefore significantly underestimate their total mass. These prominences may also escape detection totally if they were not in view during the (often brief) observing window.
%As a result, a lack of observed prominences may underestimate their occurrence \citep{2014MNRAS.443..898L,2017IAUS..328..198K}. 
For some stars, however, such as LQ Lup, even although the star has a low inclination angle and so is observed almost pole-on, if the  co-rotation radius is close enough to the surface that the prominence system is seen in emission, then the entire prominence system may be detected \citep{2000MNRAS.316..699D}. 

A promising approach is to search for the velocity shifts associated with the destabilisation of prominences \citep{2014MNRAS.443..898L,2017IAUS..328..198K,2019A&A...623A..49V}. For slingshot prominences, destabilisation can lead to ejection if the material is supported beyond the co-rotation radius, or draining back to the surface if it has cooled below the co-rotation radius and so lies in the ``hydrostatic'' regime \citep{2019MNRAS.482.2853J}. For solar-like prominences (whose support and destabilisation may be due to different processes) ejection is of course possible from below the co-rotation radius. Line asymmetries alone cannot distinguish between these two types of prominences. \citet{2019A&A...623A..49V} present an analysis of velocity shifts detected in a large number of cool stars. The majority of these shifts imply projected velocities below the escape speed. These disruptions will not lead to ejection, and hence will not influence the mass or angular momentum loss rates.

Our estimates of mass loss rates in prominences for AB Dor show that they can form a significant (though not dominant) contribution to the stellar wind. Year-to-year increases in the base density (produced by changes in the surface field strength) and the inclination of the dipole component of the field can increase the mass loss rates in both prominences and the wind by a factor of 100. This may explain the large scatter in mass loss rates predicted by the measurements of \citet{2004LRSP....1....2W}. These values are consistent with the wind mass loss rates for AB Dor predicted by \citet{2010ApJ...721...80C} using a 3D wind model based on the surface magnetograms from Dec 2007. In the case of the Sun, the wind mass loss rate varies by only a factor of two over the solar cycle \citep{2006ApJ...653..708W}. \citet{2011MNRAS.417.2592C} suggests that this is due to the relative constancy of the Sun's open magnetic flux, compared to the larger variations in the closed flux that drives the factor of 10-100 variation in the solar X-ray luminosity over the cycle. For AB Dor, as in other stars however,  the combined effect of fluctuations in the strength of the dipole field and the latitude of the dipole axis lead to a significant variation in the mass loss rates from one epoch to the next. To date around 20 stars have observations of their magnetic field geometries made over several epochs (see, for example, \citet{ 2018A&A...620L..11B} and \citet{2017MNRAS.471L..96J} and references therein). Studies showing a full polarity switch of the dipole axis such as \citet{2013MNRAS.435.1451F,2018A&A...620L..11B} may therefore allow us to quantify mass loss variations in other stars.

A simple analysis of a magnetised wind \citep{1967ApJ...148..217W} reveals that the accumulated torques exchanged as the outflowing plasma interacts with the magnetic field are equivalent to those removed by plasma ejected directly from the Alfv\'en radius $r_A$, giving a loss of specific angular momentum $L=r_A^2 \omega_\star$. In the case of the ejection of prominences, however, we have assumed that they remove the specific angular momentum $L=r_k^2 \omega_\star$ that they possessed when they were formed at the co-rotation radius $r_k$. Since $r_A>r_k$, even if the prominence-bearing flux tubes had the same surface area coverage as the wind-bearing flux tubes, the wind would remove more angular momentum. Our estimate of the torques exerted by prominences when they are ejected are however lower limits, since we neglect the torque they exert after they lose equilibrium, but while they are still interacting with the stellar wind.  Including this contribution would enhance the angular momentum loss rates predicted by current braking laws based solely on the wind, such as  \citep{2012ApJ...754L..26M,2015ApJ...798..116R,2018ApJ...854...78F}. This may be especially important in the young rapid rotators where prominence formation and ejection is most frequent.  

In addition to providing a lower limit to the torques, prominence ejection may also provide a lower limit to the rate of CME ejection. \citet{2013AN....334...77A} considered the role of CME ejection in the angular momentum loss from such young stars and determined that if the mass loss rate in CMEs was greater than $10^{-10}$M$_\odot$yr$^{-1}$ then they could significantly influence the star's rotational history. As noted by \citet{2013ApJ...764..170D} a simple extrapolation from the relationship between solar CME kinetic energies and X-ray fluence leads to unphysically large energies for stellar CMEs. One possible solution is that the strong magnetic field in such stars may suppress the ejection of CMEs (even if a flare is observed) \citep{2018ApJ...862...93A}.

Even in the case of the Sun, where CMEs may be imaged and detected with in situ measurements, the relationship between flares, prominence eruption and CME ejection is still unclear. We do not know the rate of CME ejection for other stars, although observational studies are ongoing \citep{2018ApJ...856...39C}.  Using an empirical approach, \citet{2017MNRAS.472..876O} also find extremely high mass loss rates, especially for the most active young stars. \citet{2017ApJ...840..114C} has generalised to other stars the solar power-law relationship between magnetic filling factor and CME kinetic energy flux. This predicts that for solar-like stars with ages less than 1 Gyr, the mass loss rate in CMEs exceeds that of the stellar wind. A better understanding of the dynamics of the large ``slingshot'' prominences, whose ejection may be associated with large CMEs, may provide more insight into this form of stellar mass loss.

\section{Summary and Conclusions}

We present a new method of modelling the observational signatures of ``slingshot prominences" - clouds of cool plasma trapped within the coronae of rapidly-rotating stars. We use maps of the surface magnetic fields of the star as inputs for a model of the 3D structure of both the coronal magnetic field and the coronal gas. Within this field we determine the sites of stable mechanical equilibria where material that has cooled could be supported. This cool gas not only outlines the structure of the magnetic field, but it scatters photospheric photons out of the line of sight, producing transient absorption features in  H-$\alpha$. The drift rates and observation phases at which these transient features appear in the stacked  H-$\alpha$ spectra therefore reveal the distribution of absorbing material in the corona.

From our model of the distribution of cool prominences in the stellar corona, we produce synthetic H-$\alpha$ stacked spectra and compare with the observed spectra. As an example we consider AB Dor, a young rapid rotator whose prominences and surface magnetic field have been well observed approximately annually between 1995 and 2007. Our results fall into four broad categories:
\begin{itemize}
    \item We have explored the synthetic H-$\alpha$ spectra produced by different types of field extrapolation. We show that a potential field extrapolation supports a distribution of prominences clustered at the co-rotation radius (which for AB Dor is at a radius of 2.7 R$_\star$), consistent with observation. The corresponding synthetic H-$\alpha$ absorption traces have drift rates (which are proportional to the prominence distance from the rotation axis) that agree well with what is observed. By comparison, a non-potential field extrapolation supports only prominences at much lower heights and produces absorption transients with much lower drift rates. We therefore conclude that any volume currents present in the coronal magnetic field must have decayed sufficiently rapidly with height that they are not supported at the co-rotation radius. The strong toroidal fields detected at the surface must have decayed with height sufficiently rapidly that they do not influence the locations of prominence support. 

\item We have investigated the distribution of prominence material around the star. The field at large heights is dominated by the dipole term, and hence prominence plasma preferentially accumulates at the two longitudes where the magnetic equator of this dipole intersects the rotational equator. For many (although not all) inclinations of the rotational axis to the observer, this will produce two clumps of absorption in the dynamic H-$\alpha$ spectra, symmetrically placed on either side of the rotation phase of the dipole axis.  If the dipole axis is close to the rotation axis, a torus of cool material may be supported in the plane of the rotational equator.

\item We have also explored the factors that determine the total mass that can be supported in prominences. We find that this depends on $B^2$ and also on the inclination of the dipole axis to the rotation axis. At low inclinations, the summits of the largest loops (where the magnetic field can provide support against centrifugal ejection) lie close to the rotational equator, where rotational support is greatest. As a result, magnetic fields at these low inclinations support the greatest mass. At higher inclinations, only a subset of loop summits lie close to the rotational equator, and so the total mass that can be supported is reduced. We conclude that changes in the field strength and geometry may have a significant effect on the prominence mass. For AB Dor, over the period of observations, this results in a variation of (0.1-30) $\times 10^{14}$kg in the total mass that can be supported, which compares well with the range of observed masses of (2-6)  $\times 10^{14}$kg.  Of this total mass, not all will transit the stellar disk from the point of view of the observer. For AB Dor, whose rotation axis is inclined at 60$^\circ$ to the observer, typically $<50\%$  percent of the mass can be detected as transient H-$\alpha$ absorption features. For cases where the dipole axis is almost aligned with the magnetic axis, this fraction can be as low as $2\%$ Thus estimates of the mass lost when these prominences are ejected may be significant underestimates.

\item Finally, we note that AB Dor has a coronal temperature (based on its X-ray flux) that is sufficiently high that the upflows that would form its prominences would be supersonic by the time they would reach the prominence formation site \citep{2019MNRAS.482.2853J}. As a result, the surface would be unable to respond to the growing mass in the prominence and would continue to drive an upflow. Prominences would form and be ejected (when the maximum supportable mass was reached) in a limit-cycle. We have calculated the rates of loss of mass and angular momentum associated with this form of repeated prominence ejection, which for AB Dor over the 11 years of observation vary between (0.4-24) $\times 10^{-14}$ M$_\odot$/yr and (0.2-20) $\times 10^{32}$erg respectively. These rates are less than, but within the range of, those predicted by our wind model, which are (0.8-63) $\times 10^{-14}$M$_\odot$/yr and (4-164)$\times 10^{32}$erg respectively. This suggests that while AB Dor's prominences might not be dominating the angular momentum budget, their ejection will enhance the stellar wind. 
\end{itemize}

\section*{Acknowledgements}

MJ and ACC acknowledge support from STFC consolidated grant number ST/R000824/1. JFD acknowledges funding from the European Research Council (ERC) under the H2020 research and innovation programme (grant agreement 740651 NewWorlds). 

%%%%%%%%%%%%%%%%%%%%%%%%%%%%%%%%%%%%%%%%%%%%%%%%%%



\bibliographystyle{mnras}
\bibliography{Halpha} % if your bibtex file is called example.bib

\bsp	% typesetting comment
\label{lastpage}
\end{document}